\documentclass[11pt]{article}
\usepackage[english]{babel}
\usepackage{amssymb,amstext,amsfonts}
\usepackage{amsmath}
\usepackage{setspace} 

\csname @addtoreset\endcsname{equation}{section}

\newcommand{\ft}[2]{{\textstyle\frac{#1}{#2}}}

\def\Re{\mathop{\rm Re}\nolimits}
\def\Im{\mathop{\rm Im}\nolimits}
\def\trace{\mathop{\rm Tr}\nolimits}

\def\rmi{{\rm i}}
\def\rmd{{\rm d}}

\newsavebox{\uuunit}
\sbox{\uuunit}
    {\setlength{\unitlength}{0.825em}
     \begin{picture}(0.6,0.7)
        \thinlines
        \put(0,0){\line(1,0){0.5}}
        \put(0.15,0){\line(0,1){0.7}}
        \put(0.35,0){\line(0,1){0.8}}
       \multiput(0.3,0.8)(-0.04,-0.02){12}{\rule{0.5pt}{0.5pt}}
     \end {picture}}
\newcommand {\unity}{\mathord{\!\usebox{\uuunit}}}

\usepackage{ifpdf}
\newif\ifpdf
\ifx\pdfoutput\undefined
   \pdffalse
   \usepackage{cite}
 \else
   \pdfoutput=1
   \pdftrue
  \usepackage[pdftex]{hyperref}
\fi

\begin{document}

\onehalfspacing 


\begin{titlepage}
\begin{flushright}
CERN-PH-TH/2008-172\\
KUL-TF-08/15\\
MPP-2008-97\\
arXiv:0808.2130
\end{flushright}
\vspace{.5cm}
\begin{center}
\baselineskip=16pt {\LARGE    \textbf{Electric/Magnetic Duality for Chiral Gauge}\\
\vskip5mm \textbf{Theories with Anomaly Cancellation}
}\\
\vfill
{\large Jan De Rydt $^{1,2}$, Torsten T. Schmidt $^{3}$, Mario Trigiante $^4$, \\
\vskip 0.2cm
 Antoine Van Proeyen $^1$ and Marco Zagermann $^{3}$ 
  } \\
\vfill
{\small $^1$ Instituut voor Theoretische Fysica, Katholieke Universiteit Leuven,\\
       Celestijnenlaan 200D B-3001 Leuven, Belgium
      \\[2mm]
      $^2$ Physics Department,Theory Unit, CERN,\\
CH 1211, Geneva 23, Switzerland\\[2mm]
      $^3$ Max-Planck-Institut f{\"u}r Physik, F{\"o}hringer Ring 6,  \\ 80805 M{\"u}nchen, Germany
\\[2mm]
$^4$ Dipartimento di Fisica \& INFN, Sezione di Torino, Politecnico di
Torino\\
C. so Duca degli Abruzzi, 24, I-10129 Torino, Italy\\
 \vspace{6pt}
 }
\end{center}
\vfill
\begin{center}
{\bf Abstract}
\end{center}
{\small We show that 4D gauge theories with Green-Schwarz anomaly
cancellation and possible generalized Chern-Simons terms admit a
formulation that is manifestly covariant with respect to
electric/magnetic duality transformations. This generalizes previous work
on the symplectically covariant formulation of \emph{anomaly-free} gauge
theories as they typically occur in extended supergravity, and now also
includes general theories with (pseudo-)anomalous gauge interactions as
they may occur in global or local $\mathcal{N}=1$ supersymmetry. This
generalization is achieved by relaxing the linear constraint on the
embedding tensor so as to allow for a symmetric 3-tensor related to
electric and/or magnetic quantum anomalies in these theories. Apart from
electric and magnetic gauge fields, the resulting Lagrangians also
feature two-form fields and can accommodate various unusual duality
frames as they often appear, e.g., in string compactifications with
background fluxes. } \vfill

\hrule width 3.cm \vspace{2mm}{\footnotesize \noindent e-mails:
\{Jan.DeRydt, Antoine.VanProeyen\}@fys.kuleuven.be, \{schto,
zagerman\}@mppmu.mpg.de,\\\phantom{e-mails: } mario.trigiante@to.infn.it
}
\end{titlepage}
\addtocounter{page}{1}
 \tableofcontents{}
\newpage




\normalsize
\section{Introduction}
In field theories with chiral gauge interactions, the requirement of
anomaly-freedom imposes a number of nontrivial constraints on the
possible gauge quantum numbers of the chiral fermions. The strongest
requirements are obtained if one demands that all anomalous one-loop
diagrams due to chiral fermions simply add up to zero.

These constraints on the fermionic spectrum can be somewhat relaxed if
some of the anomalous one-loop contributions are instead cancelled by
\emph{classical} gauge-variances of certain terms in the tree-level
action. The prime example for this  is the Green-Schwarz mechanism
\cite{Green:1984sg}. In its four-dimensional incarnation, it uses the
gauge variance of Peccei-Quinn terms of the form $a\mathcal{F}\wedge
\mathcal{F}$, with $a(x)$ being an axionic scalar field and $\mathcal{F}$
some vector field strengths, under gauged shift symmetries of the form
$a(x)\rightarrow a(x) + c \Lambda(x)$, where $\Lambda(x)$ is the local
gauge parameter and $c$ a constant. Gauge variances of this form may
cancel mixed Abelian/non-Abelian as well as cubic Abelian gauge anomalies
in the quantum effective action. The Abelian gauge bosons that implement
the gauged shift symmetries of the axions via St{\"u}ckelberg-type gauge
couplings correspond to the anomalous Abelian gauge groups and gain a
mass due to their St{\"u}ckelberg couplings. If their masses are low enough,
these pseudo-anomalous gauge bosons might be observable and could
possibly play the r{\^o}le of a particular type of $Z^{\prime}$-boson. The
phenomenology of such St{\"u}ckelberg $Z^{\prime}$-extensions of the Standard
Model was studied in various works \cite{Ghilencea:2002da,Kors:2004dx,
Kors:2004ri, Kors:2005uz, Feldman:2006wd, Feldman:2006ce, Feldman:2007wj,
Cheung:2007uu, Langacker:2008yv}, which were in part inspired by
intersecting brane models in type II orientifolds, where the operation of
a 4D Green-Schwarz mechanism is quite generic
\cite{Aldazabal:2000dg}.\footnote{For more details on intersecting brane
models, see, e.g. the reviews
\cite{Angelantonj:2002ct,Uranga:2003pz,MarchesanoBuznego:2003hp,Lust:2004ks,Blumenhagen:2005mu,Blumenhagen:2006ci}
and references therein.}

In \cite{Anastasopoulos:2006cz,Anastasopoulos:2007qm,Coriano':2005js},
however, it has recently been pointed out that in these orientifold
compactifications, the Green-Schwarz mechanism is often not sufficient to
cancel all quantum anomalies.\footnote{See also
\cite{Kumar:2007zza,Armillis:2007wb,Armillis:2007tb,Antoniadis:2007sp,Harvey:2007ca}.}
In particular, the cancellation of mixed Abelian anomalies between
anomalous and non-anomalous Abelian factors in general needs an
additional ingredient, so-called generalized Chern-Simons terms (GCS
terms), in the classical action. GCS terms are of the schematic form
$A\wedge A\wedge dA$ and $A \wedge A \wedge A \wedge A$, where the vector
fields $A$ are not all the same. It is quite obvious that GCS terms are
not gauge invariant, and it is precisely this gauge variance that can be
used in some cases to cancel possible left-over gauge variances from
quantum anomalies and Peccei-Quinn terms. Interestingly, these GCS terms
indeed do occur quite generically in the above-mentioned orientifold
compactifications \cite{Anastasopoulos:2006cz,Coriano':2005js}.
Phenomenologically, they provide extra trilinear (and quartic) couplings
between anomalous and non-anomalous gauge bosons, which, given a low
St{\"u}ckelberg mass scale, may lead to $Z^{\prime}$-bosons with possibly
observable new characteristic signals
\cite{Anastasopoulos:2006cz,Anastasopoulos:2007qm,Coriano':2005js}.

In \cite{DeRydt:2007vg}, it is shown how models with all three ingredients (each of which individually breaks gauge symmetry):
\begin{enumerate}
 \item anomalous fermionic spectra,
\item Peccei-Quinn terms with gauged axionic shift symmetries,
\item generalized Chern-Simons terms,
\end{enumerate}
can be compatible with global and local $\mathcal{N}=1$ supersymmetry.
This compatibility is non-trivial, because a violation of gauge
symmetries usually also triggers a violation of the on-shell
supersymmetry, as is best seen by recalling that in the Wess-Zumino gauge
the preserved supersymmetry is a combination of the original superspace
supersymmetry and a gauge transformation. Due to the presence of the
quantum gauge anomalies, one therefore also has to take into account the
corresponding supersymmetry anomalies of the quantum effective action, as
they have been determined by Brandt for $\mathcal{N}=1$ supergravity in
\cite{Brandt:1993vd,Brandt:1997au}. A recent application of the theories
studied in \cite{DeRydt:2007vg} to globally supersymmetric models with
interesting phenomenology appeared in \cite{Anastasopoulos:2008jt}.

While in \cite{Anastasopoulos:2006cz,DeRydt:2007vg} the general interplay
of all the above three ingredients is discussed, it should be emphasized
that not all three ingredients necessarily need to be present in a gauge
invariant theory. This is obvious from the original St{\"u}ckelberg
$Z^{\prime}$-models \cite{Kors:2004dx, Kors:2004ri, Kors:2005uz,
Feldman:2006wd, Feldman:2006ce, Feldman:2007wj, Cheung:2007uu,
Langacker:2008yv}, which do not have GCS terms. However, one can also
construct purely classical theories, in which only the last two
ingredients (ii) and (iii), i.e.  the gauged shift symmetries and the GCS
terms, are present and the fermionic spectrum is either absent or
non-anomalous. In fact, it was in such a context that GCS terms were
first discussed in the literature. More concretely, their possibility was
first discovered in extended gauged supergravity theories
\cite{deWit:1985px}, which are automatically free of quantum anomalies
due to the incompatibility of chiral gauge interactions with extended 4D
supersymmetry. The ensuing papers
\cite{deWit:1987ph,deWit:2002vt,deWit:2003hr,Schon:2006kz,Derendinger:2007xp,deWit:2007mt,D'Auria:2003jk,Angelantonj:2003rq,Gunaydin:2005bf,Aharony:2008rx}
likewise remained focused on -- or were inspired by -- the structures
found in extended supergravity. Recently, axionic gaugings and GCS terms
were also considered in the context of global $\mathcal{N}=1$
supersymmetry in \cite{Andrianopoli:2004sv}. In all these cases, the
absence of quantum anomalies restricts the form of the possible gauged
axionic shift symmetries.

Another very important example in this context is the work
\cite{deWit:2005ub}, which combines \emph{classically} gauge invariant
local Lagrangians that may also include Peccei-Quinn and GCS terms with
the concept of electric/magnetic duality transformations. In four
spacetime dimensions, a field theory with $n$ Abelian vector potentials
and no charged matter fields admits reparametrizations in the form of
electic/magnetic duality transformations. Those transformations that
leave the set of field equations and Bianchi identities invariant are the
rigid (or global) symmetries of the theory and form the global symmetry
group $G_{\rm rigid}$. In section \ref{ss:symplcovgauging}, we will
discuss how, in general, $G_{\rm rigid}$ is contained in the direct
product of the symplectic duality transformations that act on the vector
fields and the isometry group of the scalar manifold of the chiral
multiplets: $G_{\rm rigid} \subseteq Sp(2n,\mathbb{R})\times
\textrm{Iso}(\mathcal{M}_{\rm scalar})$.

Note, however, that the Lagrangians that encode the field equations are
in general not invariant under such rigid symmetry transformations, as
the latter may involve nontrivial mixing of field equations and Bianchi
identities. Moreover, the fields before and after a symmetry
transformation are, in general, not related by a \emph{local} field
transformation.

In order to gauge a rigid symmetry in the standard way (i.e., in order to
introduce charges for some of the fields), one needs to be able to go to
a symplectic duality frame in which the symmetry leaves the action
invariant. This automatically implies that the symmetry is also
implemented by \emph{local} field transformations. This would then allow
the introduction of minimal couplings and covariant field strengths for
the electric vector potentials in the Lagrangian in the usual way. This
standard procedure obviously singles out certain duality frames and
breaks the original duality covariance.

In \cite{deWit:2005ub}, it was shown how one can nevertheless reformulate
4D gauge theories in such a way as to maintain, formally, the full
duality covariance of the original ungauged theory. In order to do so,
the authors consider electric and magnetic gauge potentials
$(A_{\mu}{}^{\Lambda}, A_{\mu\, \Lambda})$ $(\Lambda=1,\ldots, n)$ at the
same time and combine them into a $2n$-plet, $A_{\mu}{}^{M}$
$(M=1,\ldots,2n)$ of vector potentials. Introducing then also a set of
antisymmetric tensor fields, an intricate system of gauge invariances can
be implemented, which ensures that the number of propagating degrees of
freedom is the same as before. The coupling of the electric and magnetic
vector potentials to charged fields is then encoded in the so-called
embedding tensor
$\Theta_{M}{}^{\alpha}=(\Theta_{\Lambda}{}^{\alpha},\Theta^{\Lambda \,
\alpha})$,   which enters the covariant derivatives of matter fields,
$\phi$, schematically,
\begin{equation}
(\partial_{\mu}-A_{\mu}{}^{M}\Theta_{M}{}^{\alpha} \delta
_{\alpha})\phi\,. \label{covderTheta}
\end{equation}
Here, $\alpha=1,\ldots, \dim (G_{\rm rigid})$ labels the generators of
the rigid symmetry group, $G_{\rm rigid}$, acting as $\delta
_{\alpha}\phi $ on the matter fields. In general, the gauge group also
acts on the vector fields via $(2n\times 2n)$-matrices,
\begin{equation}
  (X_M)_N{}^P\equiv X_{MN}{}^P\equiv\Theta_M{}^\alpha
(t_{\alpha})_{N}{}^P\,,
 \label{defXMNP}
\end{equation}
where the $(t_{\alpha})_{N}{}^P$ are in the fundamental representation of
$Sp(2n,\mathbb{R})$.

The embedding tensor has to satisfy a quadratic constraint in order to
ensure the closure of the gauge algebra inside the algebra of $G_{\rm
rigid}$. In \cite{deWit:2005ub}, this fundamental constraint is
supplemented by one additional constraint linear in the embedding tensor,
which can be written in terms of the above-mentioned tensor
$X_{MN}{}^{P}$, as\footnote{This constraint was considered in
\cite{deWit:2005ub} for general $\mathcal{N}$ and in particular for
$\mathcal{N}=1$ gauged supergravity and generalizes an analogous
condition originally found in \cite{deWit:1985px}. In the context of
rigid $\mathcal{N}=1$ supersymmetry, its electric version already
appeared in \cite{Andrianopoli:2004sv}.}
\begin{equation}
X_{(MN}{}^{Q}\Omega_{P)Q}=0\,, \label{repcon}
\end{equation}
where $\Omega_{PQ}$ is the symplectic metric of $Sp(2n,\mathbb{R})$. This
constraint is sometimes called the ``representation constraint'', as it
suppresses a representation of the rigid symmetry group in the tensor
$X_{MN}{}^{P}$. Together with the quadratic constraint, it ensures mutual
locality of all physical fields that are present in the
action.\footnote{A subtlety arises for generators $\delta_{\alpha}$ that
have a trivial action on the vector fields, i.e.,
$(t_{\alpha})_{M}{}^{N}=0$. In that case the mutual locality of the
corresponding electric/magnetic components of the embedding tensor should
be imposed as an independent quadratic constraint.} The full physical
meaning of this additional constraint, however, always remained a bit
obscure, and was inferred in \cite{deWit:2005ub} from
 identities that are known to be valid in $\mathcal{N}=8$ or $\mathcal{N}=2$ supergravity.

In this paper, we propose a physical interpretation of this representation constraint and recognize it as the condition for the \emph{absence of quantum anomalies}. Quantum anomalies are automatically absent in extended 4D supergravity theories, and so it is no surprise, that the internal consistency of $\mathcal{N}=8$ or $\mathcal{N}=2$ supergravity always hinted at
the validity of the constraint (\ref{repcon}).

We then go one step further and show that if quantum anomalies
proportional to a constant, totally symmetric tensor,\footnote{The tensor
$d_{MNP}$ is the one that defines the consistent anomaly in the form
given in equation (\ref{finalanomaly}). As the gauge symmetry in the
matter sector is implemented by minimal couplings to the gauge potentials
dressed with an embedding tensor, as can be seen from
(\ref{covderTheta}), the tensor $d_{MNP}$ must be of the form
(\ref{relrepcon}).} $d_{MNP}$, are present, the representation constraint
(\ref{repcon}) has to be relaxed to
 \begin{equation}
 X_{(MN}{}^{Q}\Omega_{P)Q}=d_{MNP}\,,\qquad \mbox{with}\qquad
 d_{MNP}=\Theta_M{}^\alpha \Theta _N{}^\beta \Theta_P{}^\gamma d_{\alpha \beta \gamma}\,,
 \label{relrepcon}
 \end{equation}
to allow for a gauge invariant quantum effective action. Here $d_{\alpha
\beta \gamma }$ is a symmetric tensor that will be defined by the
anomalies. We show explicitly how the framework of \cite{deWit:2005ub}
has to be modified in such a situation and that the resulting gauge
variance of the classical Lagrangian precisely gives the negative of the
consistent quantum anomaly encoded in $d_{MNP}$.

Our work can thus be viewed as a generalization of \cite{deWit:2005ub} to
theories with quantum anomalies or, equivalently, as the covariantization
of \cite{Anastasopoulos:2006cz,DeRydt:2007vg} with respect to
electric/magnetic duality transformations, and includes situations in
which pseudo-anomalous gauge interactions are mediated by magnetic vector
potentials. While already interesting in itself, our results promise to
be very useful for the description of flux compactifications with chiral
fermionic spectra, as e.g. in intersecting brane models on orientifolds
with fluxes, because flux compactifications often give 4D theories which
appear naturally in unusual duality frames and contain two-form fields.

The outline of this paper is as follows. In section \ref{ss:Anomalies},
we briefly recapitulate the results of \cite{DeRydt:2007vg}, adapted to
the notation of \cite{deWit:2005ub}. Section \ref{ss:embedding} then
gives the symplectically covariant framework of \cite{deWit:2005ub} in a
more general treatment without using the representation constraint
(\ref{repcon}). In section \ref{ss:symmetric} we show how the formalism
of \cite{deWit:2005ub} has to be modified in order to accommodate quantum
anomalies involving the relaxed representation constraint
(\ref{relrepcon}). We flesh out our results with a simple nontrivial
example in section \ref{ss:example} and conclude in
section~\ref{ss:Conclusions}.




\section{Anomalies, generalized Chern-Simons terms and gauged shift symmetries in $\mathcal{N}=1$ supersymmetry}

\label{ss:Anomalies}

In this section, we summarize the results of \cite{DeRydt:2007vg} which will later motivate our proposed generalization (\ref{relrepcon}) of the original constraint (\ref{repcon}).

In a generic low energy effective field theory, the kinetic and the theta
angle terms of vector fields, $A_{\mu}{}^{\Lambda}$, appear with scalar
field dependent coefficients\footnote{To compare notations between this
paper, ref. \cite{DeRydt:2007vg} and ref. \cite{deWit:2005ub}, note that
the vector fields were denoted as $W_\mu {}^A$ in \cite{DeRydt:2007vg},
and are here and in \cite{deWit:2005ub} denoted as $A_\mu{} ^\Lambda$
(upper greek letters are electric indices). In \cite{DeRydt:2007vg}, the
kinetic matrix for the vector multiplets is, as in most of the ${\cal
N}=1$ literature,  denoted as $f_{AB}$, which corresponds to $-\rmi {\cal
N}^*_{\Lambda\Sigma}$ in this paper. The structure constants
$f_{AB}{}^{C}$ of  \cite{DeRydt:2007vg} correspond to the
$X_{\Lambda\Sigma}{}^{\Omega}=f_{\Lambda\Sigma}{}^{\Omega}$ here, and the
axionic shift tensors $C_{AB,C}$ of \cite{DeRydt:2007vg} are now called
$X_{\Lambda\Sigma\Omega}=X_{\Lambda(\Sigma\Omega)}=C_{\Sigma\Omega,\Lambda}$.
To compare formulae of \cite{deWit:2005ub} to those here and in
\cite{DeRydt:2007vg}, the Levi-Civita symbol $\varepsilon ^{\mu \nu \rho
\sigma }$ appears in covariant equations with opposite sign (but
$\varepsilon _{0123}=1$ is valid in both cases  due to another
orientation of the spacetime directions).},
\begin{eqnarray}
{\cal L}_{\text{g.k.}}
& = &
   \frac{1}{4} e \mathcal{I}_{\Lambda\Sigma}(z,\bar{z}) \mathcal{F}_{\mu\nu}{}^{\Lambda} {\cal F}^{\mu \nu\Sigma}
    -\frac{1}{8}\, \mathcal{R}_{\Lambda\Sigma}(z,\bar{z})\varepsilon^{\mu\nu\rho\sigma} \mathcal{F}_{\mu\nu}{}^{\Lambda} {\cal F}_{\rho\sigma}{}^{\Sigma}\,.
 \label{Lgk}
\end{eqnarray}
Here, $\mathcal{F}_{\mu\nu}{}^{\Lambda}\equiv
2\partial_{[\mu}A_{\nu]}{}^{\Lambda}+
X_{\Sigma\Omega}{}^{\Lambda}A_{\mu}{}^{\Sigma}A_{\nu}{}^{\Omega}$ denotes
the non-Abelian field strengths with
$X_{\Sigma\Omega}{}^{\Lambda}=X_{[\Sigma\Omega]}{}^{\Lambda}$ being the
structure constants of the gauge group. We use the metric signature
$(-+++)$ and work with real $\varepsilon_{0123}=1$. As usual, $e$ denotes
the vierbein determinant. The second term in (\ref{Lgk}) is often
referred to as the Peccei-Quinn term, and the functions
$\mathcal{I}_{\Lambda\Sigma}(z,\bar{z})$ and
$\mathcal{R}_{\Lambda\Sigma}(z,\bar{z})$ depend nontrivially on the
scalar fields, $z^{i}$, of the theory. One can combine these functions to
a complex function
$\mathcal{N}_{\Lambda\Sigma}(z,\bar{z})=\mathcal{R}_{\Lambda\Sigma}(z,\bar{z})+\rmi\mathcal{I}_{\Lambda\Sigma}(z,\bar{z})$.
In a supersymmetric context, $\mathcal{N}_{\Lambda\Sigma}(z,\bar{z})$ has
to satisfy certain conditions, depending on the amount of supersymmetry.
In $\mathcal{N}=1$ global and local supersymmetry, which will be the
subject of the remainder of this section,
$\mathcal{N}_{\Lambda\Sigma}=\mathcal{N}_{\Lambda\Sigma}(\bar{z})$ simply
has to be antiholomorphic in the complex scalars of the chiral
multiplets.

If, under a gauge transformation with gauge parameter
$\Lambda^{\Omega}(x)$, acting on the field strengths as $\delta (\Lambda
) {\cal F}_{\mu \nu }^\Lambda =\Lambda ^\Xi {\cal F}_{\mu \nu }^\Omega
X_{\Omega \Xi }{}^\Lambda $, some of the $z^i$ transform nontrivially,
this may induce a corresponding gauge transformation of
$\mathcal{N}_{\Lambda\Sigma}(\bar{z})$. In case this transformation is of
the form of a symmetric product of two adjoint representations of the
gauge group,
 \begin{equation}
\delta(\Lambda)  \mathcal{N}_{\Lambda\Sigma}= \Lambda^{\Omega}\delta_{\Omega} \mathcal{N}_{\Lambda\Sigma}\,,\qquad \delta_{\Omega}
\mathcal{N}_{\Lambda\Sigma}=X_{\Omega\Lambda}{}^{\Gamma} \mathcal{N}_{\Sigma\Gamma} + X_{\Omega\Sigma}{}^{\Gamma}\mathcal{N}_{\Lambda\Gamma}\, , \label{conditionfinvC1}
\end{equation}
the kinetic term (\ref{Lgk}) is obviously gauge invariant. This is what
was assumed in the action of general matter-coupled supergravity in
\cite{Cremmer:1983en}.\footnote{This construction of general
matter-couplings has been reviewed in \cite{Kallosh:2000ve}. There, the
possibility (\ref{conditionfinvCb}) was already mentioned, but the extra
terms necessary for its consistency were not considered.}

If, however, one takes into account also other terms in the (quantum)
effective action, a more general transformation rule for
$\mathcal{N}_{\Lambda\Sigma}(\bar{z})$ may be allowed:
\begin{equation}
\delta_{\Omega} \mathcal{N}_{\Lambda\Sigma}  =  -X_{\Omega\Lambda\Sigma}
+X_{\Omega\Lambda}{}^{\Gamma}\mathcal{N}_{\Sigma\Gamma}
+X_{\Omega\Sigma}{}^{\Gamma}\mathcal{N}_{\Lambda\Gamma}\,.
\label{conditionfinvCb}
\end{equation}
Here, $X_{\Omega\Lambda\Sigma}$ is a constant real tensor symmetric in
the last two indices, which can be recognized as a natural generalization
in the context of symplectic duality transformations
\cite{Andrianopoli:2004sv,DeRydt:2007vg}. Closure of the gauge algebra
requires the constraint
\begin{equation}
X_{\Omega\Lambda\Sigma}X_{\Gamma\Xi}{}^{\Omega}+2 X_{\Sigma[\Xi}{}^\Omega
X_{\Gamma]\Lambda\Omega}+2X_{\Lambda[\Xi}{}^\Omega
X_{\Gamma]\Sigma\Omega}=0\,.\label{closcon}
\end{equation}
If $X_{\Omega\Lambda\Sigma}$ is non-zero, this leads to a non-gauge invariance of the
Peccei-Quinn term in  ${\cal L}_{\text{g.k.}}$:
\begin{equation}
\delta(\Lambda ) {\cal L}_{\text{g.k.}}  = \frac{1}{8}
\varepsilon^{\mu\nu\rho\sigma}\, X_{\Omega\Lambda\Sigma} \Lambda
^{\Omega}
 \mathcal{F}_{\mu\nu}{}^{\Lambda}{\cal F}_{\rho\sigma}{}^{\Sigma}\,.
 \label{delLambdaSf1}
\end{equation}
For rigid parameters, $\Lambda^{\Omega}=\mathrm{const.}$, this is just a
total derivative, but for local gauge parameters, $\Lambda^{\Omega}(x)$,
it is obviously not.

In order to  understand how this broken invariance can be restored, it is convenient to split the coefficients
$X_{\Omega\Lambda\Sigma}$  into a sum,
\begin{equation}
X_{\Omega\Lambda\Sigma}=X^{\rm(s)}_{\Omega\Lambda\Sigma}+X_{\Omega\Lambda\Sigma}^{\rm(m)}\,,\qquad X^{\rm(s)}_{\Omega\Lambda\Sigma}=
X_{(\Omega\Lambda\Sigma)}\,,\qquad X_{(\Omega\Lambda\Sigma)}^{\rm(m)}=0\,,
 \label{eq:constr1_c}
\end{equation}
where $X^{\rm(s)}_{\Omega\Lambda\Sigma}$ is completely symmetric, and
$X_{\Omega\Lambda\Sigma}^{\rm(m)}$ denotes the part of mixed symmetry.
Terms of the form (\ref{delLambdaSf1}) may then in principle be cancelled
by the following two mechanisms, or a combination thereof:

\begin{enumerate}
\item As was first realized in a similar context in $\mathcal{N}=2$ supergravity
in \cite{deWit:1985px} (see also the systematic analysis
\cite{deWit:1987ph}), the gauge variation due to a non-vanishing
mixed part, $X_{\Omega\Lambda\Sigma}^{\rm(m)}\neq 0$, may be
cancelled by adding a generalized Chern-Simons term (GCS term) that
contains a cubic and a quartic part in the vector fields,
\begin{eqnarray}
  \mathcal{L}_{\rm GCS}
& = &
\frac{1}{3} \,X^{\rm (CS)}_{\Omega\Lambda\Sigma}\,\varepsilon ^{\mu \nu \rho \sigma }
  \left(A_{\mu}{}^{\Omega}A_{\nu}{}^{\Lambda} \partial_{\rho}A_{\sigma}^{\Sigma} +\frac{3}{8}\,X_{\Gamma\Xi}{}^{\Sigma} A_{\mu}{}^{\Omega}A_{\nu}{}^{\Lambda}A_{\rho}{}^{\Gamma}A_{\sigma}{}^{\Xi}\right)\,.
 \label{SCS1}
\end{eqnarray}
This term depends on a constant tensor $X^{\rm (CS)}_{\Omega\Lambda\Sigma}$, which has the same mixed symmetry structure as $X^{\rm (m)}_{\Omega\Lambda\Sigma}$. The cancellation occurs provided the
tensors $X^{\rm (m)}_{\Omega\Lambda\Sigma}$ and $X^{\rm (CS)}_{\Omega\Lambda\Sigma}$ are, in fact,  the same. It was first shown in \cite{Andrianopoli:2004sv} that such a term can exist
in rigid $\mathcal{N}=1$ supersymmetry without quantum anomalies.

\item If the chiral fermion spectrum is anomalous under the gauge group,
the anomalous triangle diagrams lead to a non-gauge invariance of the
quantum effective action $\Gamma $ for the gauge symmetry: $\delta
(\Lambda )\Gamma =\int \rmd^4x\Lambda^\Lambda {\cal A}_\Lambda $  of
the form
\begin{equation}
  {\cal A}_\Lambda=-\frac{1}{4}\varepsilon^{\mu\nu\rho\sigma} \left[  2d_{\Omega \Sigma\Lambda} \partial_{\mu}A_{\nu}{}^\Sigma
+\left( d_{\Omega \Sigma \Gamma}X_{\Lambda \Xi}{}^\Sigma +\frac32
d_{\Omega\Sigma\Lambda}X_{\Gamma \Xi}{}^\Sigma \right) A_\mu{}^\Gamma
A_\nu{}^\Xi\right]
\partial_{\rho}A_{\sigma}{}^{\Omega}\,,\label{gaugeanom}
\end{equation}
with a symmetric\footnote{More precisely, the anomalies have a scheme
dependence. As reviewed in \cite{Anastasopoulos:2006cz} one can
choose a scheme in which the anomaly is proportional to a symmetric
$d_{\Omega\Lambda\Sigma}$. Choosing a different scheme is equivalent
to the choice of another GCS term (see item (i)). We will always work
with a renormalization scheme in which the quantum anomaly is indeed
proportional to the symmetric tensor $d_{\Omega\Lambda\Sigma}$
according to (\ref{gaugeanom}).} tensor $d_{\Omega\Lambda\Sigma}$. If
\begin{eqnarray}
X_{\Omega\Lambda\Sigma}^{\rm (s)}= d_{\Omega\Lambda\Sigma}\,,
\label{Xelectrictotsym}
\end{eqnarray}
this quantum anomaly cancels the symmetric part of
(\ref{delLambdaSf1}). This is the Green-Schwarz mechanism.
\end{enumerate}

In  \cite{DeRydt:2007vg}, it was studied to what extent a general  gauge theory
of the above type (i.e., with gauged axionic shift symmetries, GCS terms and quantum gauge anomalies) can be compatible with $\mathcal{N}=1$ supersymmetry. The results can be summarized as follows:
if one takes as one's starting point the matter-coupled supergravity Lagrangian
in eq. (5.15) of reference \cite{Kallosh:2000ve},
 an axionic shift symmetry with $X_{\Lambda\Sigma\Omega}\neq 0$ satisfying the closure condition
 (\ref{closcon}) can be gauged in a way consistent with $\mathcal{N}=1$ supersymmetry
 if
\begin{enumerate}
\item a GCS term (\ref{SCS1}) with   $X^{\rm (CS)}_{\Omega\Lambda\Sigma}=X^{\rm (m)}_{\Omega\Lambda\Sigma}$ is added,
\item an additional term bilinear in the gaugini, $\lambda^{\Sigma}(x)$, and linear in the vector fields is added\footnote{A superspace expression for the sum $\mathcal{L}_{\rm GCS}+\mathcal{L}_{\rm extra}$ is known only for the case
$X_{\Lambda\Sigma\Omega}^{\rm (s)}=0$, i.e., for the case without
quantum anomalies \cite{Andrianopoli:2004sv}.}:
\begin{equation}
\mathcal{L}_{\rm extra}= -\frac{1}{4} \rmi A_{\mu}{}^{\Omega}X_{\Omega\Lambda\Sigma}\bar{\lambda}^{\Lambda}\gamma_{5}\gamma^{\mu}\lambda^{\Sigma},
\end{equation}
\item the fermions in the chiral multiplets give rise to quantum anomalies
with $d_{\Omega\Lambda\Sigma}=X^{\rm (s)}_{\Omega\Lambda\Sigma}$. The
consistent gauge anomaly, ${\cal A}_\Lambda$ is of the form
(\ref{gaugeanom}). The exact result for the supersymmetry anomaly can be
found in \cite{Brandt:1997au} or eq. (5.8) of \cite{DeRydt:2007vg}. These
quantum anomalies precisely cancel the classical gauge and supersymmetry
variation of the new Lagrangian $\mathcal{L}_{\rm old}+\mathcal{L}_{\rm
GCS} + \mathcal{L}_{\rm extra}$, where $\mathcal{L}_{\rm old}$ denotes
the original Lagrangian of \cite{Kallosh:2000ve}.
\end{enumerate}




\section{The embedding tensor and the symplectically covariant formalism}
\label{ss:embedding}

In this section, we recapitulate the results of \cite{deWit:2005ub},
which describe a symplectically covariant formulation of (classically)
gauge invariant field theories. Correspondingly, we will assume the
absence of quantum anomalies in this section.


\subsection{Electric/magnetic duality and the conventional gauging}
 \label{ss:convgauging}
In the absence of charged fields, a gauge invariant four-dimensional
Lagrangian of $n$ Abelian vector fields $A_{\mu}{}^{\Lambda}
(\Lambda=1,\ldots,n)$ only depends on their curls
$F_{\mu\nu}{}^{\Lambda}\equiv 2\partial_{[\mu}A_{\nu]}{}^{\Lambda}$.
Defining the dual magnetic field strengths
\begin{eqnarray}
G_{\mu\nu\,\Lambda}\equiv\varepsilon_{\mu\nu\rho\sigma}\frac{\partial{\cal
L}}{\partial F_{\rho\sigma}{}^{\Lambda}}\,, \label{G}
\end{eqnarray}
the Bianchi identities and field equations read
\begin{eqnarray}
\partial_{[\mu} F_{\nu\rho]}{}^{\Lambda} & = & 0 \,,\label{b}\\
\partial_{[\mu} G_{\nu\rho]\,\Lambda} & = & 0\,. \label{eom}
\end{eqnarray}
The equations of motion (\ref{eom}) imply the existence of magnetic gauge
potentials, $A_{\mu\,\Lambda}$, via
$G_{\mu\nu\,\Lambda}=2\partial_{[\mu}A_{\nu]\Lambda}$. These magnetic
gauge potentials are related to the electric vector potentials,
$A_{\mu}{}^{\Lambda}$, by nonlocal field redefinitions. The electric
Abelian field strengths, $F_{\mu\nu}{}^{\Lambda}$, and their magnetic
duals, $G_{\mu\nu\,\Lambda}$, can be combined into a $2n$-plet,
$F_{\mu\nu}{}^{M}$, such that $F^{M}=(F^{\Lambda},G_{\Lambda})$. This
allows us to write (\ref{b}) and (\ref{eom}) in the following compact
way:
\begin{eqnarray}
\partial_{[\mu}F_{\nu\rho]}{}^{M} & = & 0\,.\label{dF}
\end{eqnarray}
Apparently, equation (\ref{dF}) is invariant under general linear transformations
\begin{equation}
F^{M}\rightarrow F^{\prime M}  =  {\cal S}^{M}{}_{N}F^{N} \,,\,\text{ where }\,
{\cal S}^{M}{}_{N}\,=\,
\begin{pmatrix}
U^{\Lambda}{}_{\Sigma}&Z^{\Lambda\Sigma}
\cr
W_{\Lambda\Sigma}&V_{\Lambda}{}^{\Sigma}
\end{pmatrix} \,,
\label{G=SG}
\end{equation}
but only for symplectic matrices ${\cal S}^{M}{}_{N}\in Sp(2n,\mathbb{R})$ a  relation of the type (\ref{G}) is possible. The admissible rotations ${\cal S}^{M}{}_{N}$ thus form the group $Sp(2n,\mathbb{R})$:
\begin{equation}
  {\cal S}^T \Omega {\cal S} = \Omega,
\end{equation}
with the symplectic metric, $\Omega_{MN}$, given by
\begin{equation}
\Omega_{MN} =\begin{pmatrix}0&\Omega _\Lambda {}^\Sigma \cr \Omega
^\Lambda {}_\Sigma &0\end{pmatrix}=
\begin{pmatrix}
0 & \delta_\Lambda ^\Sigma   \cr -\delta_\Sigma^\Lambda & 0
\end{pmatrix}\,.
\label{symplstructure}
\end{equation}
We define $\Omega^{MN}$ via $\Omega^{MN}\Omega_{NP}=-\delta^{M}{}_{P}$.
Note that the components of $\Omega^{MN}$ should not be written as
$\Omega ^\Lambda {}_\Sigma $ etc., as these are different from
(\ref{symplstructure}).

Starting with a kinetic Lagrangian of the form (\ref{Lgk}),
an electric/magnetic duality transformation leads to a new Lagrangian, ${\cal L}'(F')$, which is of a similar form,
but with a new gauge kinetic function
\begin{eqnarray}
{\cal N}_{\Lambda\Sigma} & \rightarrow & {\cal
N}'_{\Lambda\Sigma}=(V{\cal N}+ W)_{\Lambda\Omega}\big[(U+Z{\cal
N})^{-1}\big]^{\Omega}{}_{\Sigma}\,. \label{rot:calN}
\end{eqnarray}

The subset of $Sp(2n,\mathbb{R})$ symmetries (of field equations and
Bianchi identities) for which the Lagrangian remains unchanged in the
sense that ${\cal L}'(F'(F))={\cal L}(F)$ and  (\ref{rot:calN}) is
implemented by transformations of the fields on which ${\cal N}$ depends,
are \emph{invariances} of the action. In a different duality frame, the
Lagrangian might have a different set of invariances.

{}From the spacetime point of view, these are all rigid (``global'')
symmetries. Sometimes these global symmetries can be turned into local
(``gauge'') symmetries. For the conventional gaugings one has to restrict
to the transformations that leave the Lagrangian invariant, which implies
that $Z^{\Lambda\Sigma}$ in the matrices ${\cal S}^{M}{}_{N}$ of
(\ref{G=SG}) has to vanish. In the context of symplectically covariant
gaugings \cite{deWit:2005ub}, however, this restriction can be lifted,
and we will come back to these in section \ref{ss:symplcovgauging}. The
standard way to perform a gauging of a symmetry of interest is therefore
to first switch to a symplectic duality frame in which the symmetries of
interest act on
$F_{\mu\nu}{}^{M}=(F_{\mu\nu}{}^{\Lambda},G_{\mu\nu\,\Lambda})$ by lower
block triangular matrices (i.e. those with $Z=0$) such that they become
(as rigid symmetries) invariances of the action.

The gauging requires the introduction of gauge covariant derivatives and
field strengths and can be implemented solely with the electric vector
fields $A_{\mu}{}^{\Omega}$ and the corresponding electric gauge
parameters $\Lambda^{\Omega}$. The gaugeable symplectic transformation,
${\cal S}$, must be of the infinitesimal form
\begin{equation}
  {\cal S}^M{}_N=\delta^M{}_N  -\Lambda ^\Omega {\cal S}_\Omega
  {}^M{}_N\,.
  \label{SSigma}
\end{equation}
According to our definition (\ref{G=SG}), these infinitesimal symplectic
transformations act on the field strengths by multiplication with the
matrices $\mathcal{S}_{\Lambda}{}^{M}{}_{N}$ from the left. Following the
conventions of \cite{deWit:2005ub}, however, we will use matrices
$X_{\Omega M}{}^{N}$ to describe the infinitesimal symplectic action via
multiplication from the right:
\begin{eqnarray}
  \delta F_{\mu \nu }{}^M&=&F'_{\mu \nu }{}^M-F_{\mu \nu }{}^M=-\Lambda ^\Omega F_{\mu \nu
  }{}^N X_{\Omega N}{}^M\,,\qquad \mbox{i.e.}\qquad X_{\Omega N}{}^M=S_\Omega {}^M{}_N\,.
\end{eqnarray}
For standard electric gaugings, we then have
\begin{eqnarray}
\delta\begin{pmatrix} F_{\mu \nu }^\Lambda \cr  G_{\mu \nu\,
\Lambda}\end{pmatrix}  &=&-\Lambda ^\Omega \begin{pmatrix}X_{\Omega\Xi
}{}^\Lambda  &0\cr X_{\Omega \Lambda \Xi }&X_\Omega {}^\Xi {}_\Lambda
\end{pmatrix}\begin{pmatrix}F_{\mu \nu }^\Xi  \cr  G_{\mu \nu\,\Xi} \end{pmatrix}\,,
 \label{defXS}
\end{eqnarray}
where $X_{\Omega \Sigma }{}^{\Lambda }=-X_\Omega
{}^{\Lambda}{}_{\Sigma}=f_{\Omega \Sigma }{}^{\Lambda }$ are the
structure constants of the gauge algebra,
and $X_{\Sigma\Xi\Gamma}=X_{\Sigma(\Xi\Gamma)}$ give rise to the axionic shifts mentioned in section \ref{ss:Anomalies} (compare (\ref{rot:calN}) with (\ref{conditionfinvCb}) for the particular choice of ${\cal S}$ given in (\ref{SSigma})).\\
The gauging then proceeds in the usual way by introducing covariant
derivatives $(\partial_{\mu}-A_{\mu}{}^{\Lambda}\delta _{\Lambda})$,
where the $\delta _\Lambda$ are the gauge generators in a suitable
representation of the matter fields. One also introduces covariant field
strengths and possibly GCS terms as described in section
\ref{ss:Anomalies}. As we assume the absence of quantum anomalies in this
section, we have to require $X_{(\Lambda\Sigma\Gamma)}=0$.


\subsection{The symplectically covariant gauging \label{ss:symplcovgauging}}

We will now turn to the more general gauging of symmetries. The group
that will be gauged is a subgroup of the rigid symmetry group. What we
mean by the rigid symmetry group is a bit more subtle in ${\cal N}=1$
supergravity (or theories without supergravity) than in extended
supergravities. This is due to the fact that in extended supergravities
the vectors are supersymmetrically related to scalar fields, and
therefore their rigid symmetries are connected to the symmetries of
scalar manifolds.

In $\mathcal{N}=1$ supersymmetry, the rigid symmetry group, $G_{\rm
rigid}$, is a subset of the product of the symplectic duality
transformations that act on the vector fields and the isometry group of
the scalar manifold of the chiral multiplets: $G_{\rm rigid}\subseteq
Sp(2n,\mathbb{R})\times \textrm{Iso}(\mathcal{M}_{\rm scalar})$. The
relevant isometries are those that respect the K{\"a}hler structure (i.e.
generated by holomorphic Killing vectors) and that also leave the
superpotential invariant (in supergravity, the superpotential should
transform according to the K{\"a}hler transformations). Elements
$(g_1,\,g_2)$ of $Sp(2n,\mathbb{R}) \times \textrm{Iso}({\cal M}_{\rm
scalar})$ that are compatible with (\ref{rot:calN}) in the sense that the
symplectic action (\ref{rot:calN}) of $g_1$ on the matrix ${\cal N}$ is
induced by the isometry $g_2$ on the scalar manifold, are rigid
(``global'') symmetries provided they also leave the rest of the theory
(deriving from scalar potentials, etc.) invariant \cite{Gaillard:1981rj}.
The rigid symmetry group, $G_{\rm rigid}$, is thus a subgroup of
$Sp(2n,\mathbb{R}) \times {\rm Iso}({\cal M}_{\rm
scalar})$.\footnote{Note that this may include cases where either the
symplectic transformation $g_1$ or the isometry $g_2$ is trivial. Another
special case is when the isometry $g_2$ is non-trivial, but ${\cal N}$
does not transform under it, as happens, e.g, when ${\cal N}=\rmi \unity$
is constant. $G_{\rm rigid}$ is in general a genuine subgroup of
$Sp(2n,\mathbb{R}) \times {\rm Iso}({\cal M}_{\rm scalar})$, even in the
latter case of constant ${\cal N}$.}

The generators of $G_{\rm rigid}$ will be denoted by $\delta _{\alpha}$,
$\alpha=1,\ldots,$ dim$(G_{\rm rigid})$. These generators act on the
different fields of the theory either via Killing vectors $\delta
_\alpha=K_\alpha =K_\alpha ^i\frac{\partial }{\partial \phi ^i}$ defining
infinitesimal isometries on the scalar manifold, or with certain matrix
representations\footnote{\label{fn:structureconstants}The structure
constants defined by $[\delta _\alpha ,\delta _\beta ]=f_{\alpha \beta
}{}^\gamma \delta _\gamma $ lead for the matrices to $[t_\alpha ,t_\beta
]=-f_{\alpha \beta }{}^\gamma t_\gamma$.}, e.g. $\delta_\alpha \phi
^i=-\phi ^j(t_\alpha )_j{}^i$. 

On the field strengths
$F_{\mu\nu}{}^{M}=(F_{\mu\nu}{}^{\Lambda},\,G_{\mu\nu\,\Lambda})$, these
rigid symmetries must act by multiplication with infinitesimal symplectic
matrices\footnote{These matrices might be trivial, e.g., for Abelian
symmetry groups that only act on the scalars (and/or the fermions) and
that do not give rise to axionic shifts of the kinetic matrix
$\mathcal{N}_{\Lambda \Sigma}$.} $(t_{\alpha})_{M}{}^{P},$ i.e., we have
\begin{eqnarray}
(t_{\alpha})_{[M}{}^{P}\Omega_{N]P} & = & 0\,.
\label{symplrep}
\end{eqnarray}
In order to gauge a subgroup,  $G_{\rm local}\subset G_{\rm rigid}$, the
$2n$-dimensional vector space spanned by the vector fields
$A_{\mu}{}^{M}$ has to be projected onto the Lie algebra of $G_{\rm
local}$, which is formally done with the so-called embedding tensor
$\Theta_{M}{}^{\alpha}=(\Theta_{\Lambda}{}^{\alpha},\Theta^{\Lambda\,
\alpha})$. Equivalently, $\Theta_{M}{}^{\alpha}$ completely determines
the gauge group $G_{\rm local}$ via the decomposition of the gauge group
generators, which we will denote by $\tilde X_M$, into the generators of
the rigid invariance group $G_{\rm rigid}$:
\begin{equation}
 \tilde X_{M}\equiv\Theta_{M}{}^{\alpha}\delta _{\alpha}. \label{Xdefi}
\end{equation}
The gauge generators $\tilde X_M$ enter the gauge covariant derivatives
of matter fields,
\begin{equation}
\mathcal{D}_{\mu}  =  \partial_{\mu}-A_\mu{}^M \tilde X_M = \partial_{\mu}
-A_{\mu}{}^{\Lambda}\Theta_{\Lambda}{}^{\alpha}\delta _{\alpha}-A_{\mu\Lambda}\Theta^{\Lambda \alpha}\delta _{\alpha}\,,
\label{covderiv}
\end{equation}
where the generators $\delta _{\alpha}$ are meant to either act as
representation matrices on the fermions or as Killing vectors on the
scalar fields, as mentioned above. On the field strengths of the vector
potentials, the generators $\delta_{\alpha}$ act by multiplication with
the matrices $(t_\alpha )_N{}^P$, so that (\ref{Xdefi}) is represented by
matrices $(X_M)_N{}^P$ whose elements we denote as $X_{MN}{}^P$, see
(\ref{defXMNP}), and whose antisymmetric part in the lower indices
appears in the field strengths
\begin{equation}
\mathcal{F}_{\mu\nu}{}^{M}=2\partial_{[\mu}A_{\nu]}{}^{M}+
X_{[NP]}{}^{M}A_{\mu}{}^{N}A_{\nu}{}^{P}\,, \qquad X_{NP}{}^M=\Theta_{N}{}^{\alpha}(t_{\alpha})_P{}^M\,.
\label{defFmunuX}
\end{equation}

The symplectic property (\ref{symplrep}) implies
\begin{equation}
  X_{M[N}{}^Q\Omega _{P]Q}=0\,,\qquad X_{MQ}{}^{[N}\Omega ^{P]Q}=0\,.
 \label{fromsympl}
\end{equation}
In the remainder of this paper, the symmetrized contraction
$X_{(MN}{}^Q\Omega _{P)Q}$ will play an important r\^{o}le. We therefore
give this tensor a special name and denote it by $D_{MNP}$:
\begin{equation}
  D_{MNP}\equiv X_{(MN}{}^Q\Omega _{P)Q}\,.
 \label{defd}
\end{equation}
Note that this is really just a definition and no new constraint. Using
the definition (\ref{defd}), one can check that
\begin{eqnarray}
  &&2 X_{(MN)}{}^Q\Omega _{RQ}+X_{RM}{}^Q\Omega _{NQ}=3D_{MNR}\,,\nonumber\\
  && \mbox{i.e.}\qquad
  X_{(MN)}{}^P=\ft12\Omega ^{PR}X_{RM}{}^Q\Omega _{NQ}+\ft32D_{MNR}\Omega
  ^{RP}\,.
 \label{symmX}
\end{eqnarray}


\subsubsection{Constraints on the embedding tensor \label{ss:constraintsembedding}}

The embedding tensor $\Theta_{M}{}^{\alpha}$ has to satisfy a number of
consistency conditions. Closure of the gauge algebra and locality
require, respectively, the quadratic constraints
\begin{eqnarray}
&&\mbox{closure:}\qquad f_{\alpha\beta}{}^{\gamma}\Theta_{M}{}^{\alpha}\Theta_{N}{}^{\beta}=(t_{\alpha})_{N}{}^{P}\Theta_{M}{}^{\alpha}\Theta_{P}{}^{\gamma}   \,,\label{constraint1} \\
&&\mbox{locality:}\qquad
\Omega^{MN}\Theta_{M}{}^{\alpha}\Theta_{N}{}^{\beta}=0\hspace{.5cm}
\Leftrightarrow\hspace{.5cm}
\Theta^{\Lambda[\alpha}\Theta_{\Lambda}{}^{\beta]}  =  0\,,
\label{constraint2}
\end{eqnarray}
where $f_{\alpha\beta}{}^{\gamma}$ are the structure constants of the
rigid invariance group $G_{\rm rigid}$, see footnote
\ref{fn:structureconstants}. Another constraint, besides
(\ref{constraint1}) and (\ref{constraint2}), was inferred in
\cite{deWit:2005ub} from supersymmetry constraints in $\mathcal{N}=8$
supergravity
\begin{equation}
D_{MNR}\equiv X_{(MN}{}^{Q}\Omega_{R)Q}  =  0\,. \label{constraint3.6}
\end{equation}
This constraint eliminates some of the representations of the rigid
symmetry group and is therefore sometimes called the ``representation
constraint''. As we pointed out in the introduction, one can show that
the locality constraint is not independent of (\ref{constraint1}) and
(\ref{constraint3.6}), apart from specific cases where $(t_\alpha)_M{}^N$
has a trivial action on the vector fields.

However, we will neither use the locality constraint (\ref{constraint2})
nor the representation constraint (\ref{constraint3.6}). We will,
instead, need another constraint in section \ref{ss:topolBconstr}, whose
meaning we will discuss in section \ref{ss:symmetric}. Before coming to
that new constraint, we thus only use the closure constraint
(\ref{constraint1}). This constraint reflects the invariance of the
embedding tensor under $G_{\rm local}$ and it implies for the matrices
$X_M$ the relation
\begin{eqnarray}
[X_{M},X_{N}] & = & -X_{MN}{}^{P}\,X_{P}\,.
\label{algebra}
\end{eqnarray}
This clearly shows that the gauge group generators commute into each
other with `structure constants' given by $X_{[MN]}{}^{P}$. However, note
that $X_{MN}{}^{P}$ in general also contains a non-trivial symmetric
part, $X_{(MN)}{}^{P}$. The antisymmetry of the left hand side of
(\ref{algebra}) only requires that the contraction
$X_{(MN)}{}^{P}\Theta_{P}{}^{\alpha}$ vanishes, as is also directly
visible from (\ref{constraint1}). Therefore one has
\begin{equation}
X_{(MN)}{}^{P}\Theta_{P}{}^{\alpha}=0\qquad \rightarrow \qquad
X_{(MN)}{}^P X_{PQ}{}^R=0\,.
 \label{Xsymmcontracted0}
\end{equation}
Writing (\ref{algebra}) explicitly gives
\begin{equation}
  X_{MQ}{}^P X_{NP}{}^R - X_{NQ}{}^P X_{MP}{}^R + X_{MN}{}^PX_{PQ}{}^R=0\,.
 \label{explicitXalg}
\end{equation}
Antisymmetrizing in $[MNQ]$, we can split the second factor of each term
into the antisymmetric and symmetric part,
$X_{MN}{}^{P}=X_{[MN]}{}^{P}+X_{(MN)}{}^{P}$, and this gives a violation
of the Jacobi identity for $X_{[MN]}{}^{P}$ as
\begin{eqnarray}
&&X_{[MN]}{}^{P}X_{[QP]}{}^{R}+X_{[QM]}{}^{P}X_{[NP]}{}^{R}+X_{[NQ]}{}^{P}X_{[MP]}{}^{R} \nonumber\\
&& =-\ft13\left( X_{[MN]}{}^P X_{(QP)}{}^R+X_{[QM]}{}^{P}X_{(NP)}{}^{R}+X_{[NQ]}{}^{P}X_{(MP)}{}^{R}\right)\,.
\label{genJacobi0}
\end{eqnarray}
Other relevant consequences of (\ref{explicitXalg}) can be obtained by
(anti)symmetrizing in $MQ$. This gives, using also
(\ref{Xsymmcontracted0}), the two equations
\begin{eqnarray}
 && X_{(MQ)}{}^P X_{NP}{}^R - X_{NQ}{}^P X_{(MP)}{}^R -
 X_{NM}{}^PX_{(QP)}{}^R=0\,,\nonumber\\
&&   X_{[MQ]}{}^P X_{NP}{}^R - X_{NQ}{}^P X_{[MP]}{}^R +
   X_{NM}{}^PX_{[QP]}{}^R=0\,.
 \label{asymmetrizingexpl}
\end{eqnarray}

\subsubsection{Gauge transformations\label{ss:gaugetransf}}
The violation of the Jacobi identity (\ref{genJacobi0}) is the prize one
has to pay for the symplectically covariant treatment in which both
electric and magnetic vector potentials appear at the same time. In order
to compensate for this violation and in order to make sure that the
number of propagating degrees of freedom is the same as before, one
imposes an additional gauge invariance in addition to the usual
non-Abelian transformation
$\partial_{\mu}\Lambda^{M}+X_{[PQ]}{}^{M}A_{\mu}{}^{P}\Lambda^{Q}$ and
extends the gauge transformation of the vector potentials to
\begin{equation}
\delta A_{\mu}{}^{M}  =  {\cal
D}_{\mu}\Lambda^{M}-X_{(NP)}{}^M\Xi_\mu{}^{NP}\,, \qquad {\cal
D}_{\mu}\Lambda^{M}=\partial_{\mu}\Lambda^{M}+X_{PQ}{}^{M}A_{\mu}{}^{P}\Lambda^{Q}\,,\label{gaugevar}
\end{equation}
where we introduced the covariant derivative ${\cal D}_{\mu}\Lambda^{M}$,
and new vector-like gauge parameters $\Xi_\mu{}^{NP}$, symmetric in the
upper indices. The extra terms $X_{(PQ)}{}^{M}A_{\mu}{}^{P}\Lambda^{Q}$
and the $\Xi$-transformations contained in (\ref{gaugevar}) allow one to
gauge away the vector fields that correspond to the directions in which
the Jacobi identity is violated, i.e., directions in the kernel of the
embedding tensor (see (\ref{Xsymmcontracted0})).

It is important to notice that the modified gauge transformations
(\ref{gaugevar}) still close on the gauge fields and thus form a Lie
algebra. Indeed, if we split (\ref{gaugevar}) into two parts,
\begin{equation}
\delta A_\mu{}^M=\delta(\Lambda)A_\mu{}^M + \delta(\Xi)A_\mu{}^M\,,
\end{equation}
the commutation relations are
\begin{eqnarray}
\nonumber\left[\delta(\Lambda_1),\delta(\Lambda_2)\right]A_\mu{}^M&=&\delta(\Lambda_3)A_\mu{}^M +  \delta(\Xi_3)A_\mu{}^M,\\
  \left[\delta(\Lambda),\delta(\Xi)\right]A_\mu{}^M&=&\left[\delta(\Xi_1 ),\delta(\Xi_2)\right]A_\mu{}^M=
  0\,, \label{comm}
\end{eqnarray}
with
\begin{eqnarray}
    \Lambda_3^M&=&X_{[NP]}{}^M \Lambda_1^N \Lambda_2^P\,,\nonumber\\
   \Xi_{3\mu} {}^{PN}&=&\Lambda_1^{(P} {\cal D}_\mu \Lambda_2^{N)}-\Lambda_2^{(P} {\cal D}_\mu \Lambda_1^{N)}
   \,.
 \label{gaugeparam}
  \end{eqnarray}
To prove that the terms that are quadratic in the matrices $X_M$ in the
left-hand side of (\ref{comm}) follow this rule, one uses
(\ref{asymmetrizingexpl}).

Due to (\ref{Xsymmcontracted0}) and (\ref{gaugevar}), however, the usual
properties of the field strength
\begin{equation}
{\cal F}_{\mu\nu}{}^{M}=
2\partial_{[\mu}A_{\nu]}{}^{M}+\,X_{[PQ]}{}^{M}A_{\mu}{}^{P}A_{\nu}{}^{Q} \label{FSdef}
\end{equation}
are changed. In particular, it will no longer fulfill the Bianchi
identity, which now must be replaced by
\begin{eqnarray}
\mathcal{D}_{[\mu}\mathcal{F}_{\nu\rho]}{}^{M} & = & X_{(NP)}{}^{M}A_{[\mu}{}^{N}\mathcal{F}_{\nu\rho]}{}^{P}-\frac{1}{3} X_{(PN)}{}^{M}X_{[QR]}{}^{P}\,\,A_{[\mu}{}^{N}A_{\nu}{}^{Q}A_{\rho]}{}^{R}\,.
\label{bianchi}
\end{eqnarray}
Furthermore, $\mathcal{F}_{\mu\nu}{}^{M}$ does not transform covariantly under a gauge transformation  (\ref{gaugevar}).
Instead, we have
\begin{eqnarray}
\delta \mathcal{F}_{\mu\nu}{}^{M} & = & 2{\cal D}_{[\mu }\delta A_{\nu
]}{}^M-2X_{(PQ)}{}^{M}A_{[\mu}{}^{P}\delta A_{\nu]}{}^{Q} \nonumber\\
&=& X_{NQ}{}^{M}\ \mathcal{F}_{\mu\nu}{}^N\Lambda^{Q}-2X_{(NP)}{}^M
{\cal D}_{[\mu}\Xi_{\nu]}{}^{NP}-2X_{(PQ)}{}^{M}A_{[\mu}{}^{P}\delta
A_{\nu]}{}^{Q}\,, \label{deltaF}
\end{eqnarray}
where the covariant derivative is (both expressions are useful and
related by (\ref{asymmetrizingexpl}))
\begin{eqnarray}
X_{(NP)}{}^M {\cal D}_{\mu}\Xi_{\nu}{}^{NP}&=&\partial _\mu \left(
X_{(NP)}{}^M \Xi_{\nu}{}^{NP}\right)  + A_\mu {}^R
X_{RQ}{}^MX_{(NP)}{}^Q\Xi_{\nu}{}^{NP}\,,\nonumber\\
 {\cal D}_{\mu}\Xi_{\nu}{}^{NP}&=& \partial_{\mu}\Xi_{\nu}{}^{NP}+
X_{QR}{}^PA_{\mu}{}^{Q}\Xi_{\nu}{}^{NR}
+X_{QR}{}^NA_{\mu}{}^{Q}\Xi_{\nu}{}^{PR}\,.
 \label{DmuXi}
\end{eqnarray}
Therefore, if we want to deform the original Lagrangian (\ref{Lgk}) and
accommodate electric and magnetic gauge fields,
$\mathcal{F}_{\mu\nu}{}^{M}$ cannot be used to construct gauge-covariant
kinetic terms.

For this reason, the authors of \cite{deWit:2005ub} introduced  tensor
fields $B_{\mu\nu\, \alpha}$, later in \cite{deWit:2008ta} to be
described by $B_{\mu \nu }{}^{MN}$, symmetric in $(MN)$, and with them
modified field strengths
\begin{equation}
{\cal H}_{\mu\nu}{}^{M}  =  {\cal F}_{\mu\nu}{}^{M}+X_{(NP)}{}^M
B_{\mu\nu}{}^{NP}\,. \label{H}
\end{equation}
We will consider gauge transformations of the antisymmetric tensors of
the form
\begin{equation}
  \label{eq:B-transf-1}
    \delta B_{\mu\nu}{}^{NP} =
   2   {\cal D}_{[\mu} \Xi_{\nu]}{}{}^{NP} +2A_{[\mu}{}^{(N} \delta A_{\nu]}{}^{P)}+\Delta B_{\mu \nu }{}
   ^{NP}\,,
\end{equation}
where $\Delta B_{\mu \nu }{}^{NP}$ depends on the gauge parameter
$\Lambda ^Q$, but we do not fix it further at this point. Together with
(\ref{deltaF}), this then implies\footnote{Note that ${\cal F}_{\mu \nu
}{}^N$ in the second line of (\ref{deltaF}) can be replaced by ${\cal
H}_{\mu \nu }{}^N$ due to (\ref{Xsymmcontracted0}).\label{fn:FH}}
\begin{equation}
\delta{\cal H}_{\mu\nu}{}^{M}=X_{NQ}{}^{M}\Lambda^{Q}{\cal
H}_{\mu\nu}{}^N+ X_{(NP)}{}^M \Delta B_{\mu\nu}{}^{NP}\,.
\label{Htrafocov}
\end{equation}

\subsubsection{The kinetic Lagrangian}

The first step towards a gauge invariant action is to replace ${\cal
F}_{\mu\nu}{}^{\Lambda}$ in $\mathcal{L}_{{\rm g.k.}}$, (\ref{Lgk}), by
${\cal H}_{\mu\nu}{}^{\Lambda}$, which then yields the new kinetic
Lagrangian
\begin{equation}
\mathcal{L}_{{\rm
g.k.}}=\ft{1}{4}e\mathcal{I}_{\Lambda\Sigma}\mathcal{H}_{\mu\nu}{}^{\Lambda}\mathcal{H}^{\mu\nu\Sigma}
-\ft{1}{8}\mathcal{R}_{\Lambda\Sigma}\varepsilon^{\mu\nu\rho\sigma}\mathcal{H}_{\mu\nu}{}^{\Lambda}\mathcal{H}_{\rho\sigma}{}^{\Sigma}\,,
\label{LgkH}
\end{equation}
where again ${\cal I}_{\Lambda\Sigma}$ and ${\cal R}_{\Lambda\Sigma}$  denote, respectively,  $\Im {\cal N}_{\Lambda\Sigma}$ and $\Re {\cal N}_{\Lambda\Sigma}$.
Using
\begin{equation}
  \label{eq:def-G(H)}
  {\mathcal G}_{\mu\nu\,\Lambda} \equiv \varepsilon_{\mu\nu\rho\sigma}\frac{\partial {\cal L}}{\partial {\cal H}_{\rho\sigma}{}^\Lambda} = {\cal R}_{\Lambda\Gamma} {\cal
  H}_{\mu\nu}{}^{\Gamma} +\frac{1}{2}e\varepsilon_{\mu\nu\rho\sigma}\,
  {\cal I}_{\Lambda\Gamma}\, {\cal H}{}^{\rho\sigma\,\Gamma}\,,
\end{equation}
the Lagrangian and its transformations can be written as
\begin{eqnarray}
  {\cal L}_{{\rm g.k.}}&=&-\ft18\varepsilon^{\mu \nu \rho \sigma }{\cal H}_{\mu \nu }^\Lambda {\cal G}_{\rho \sigma \,\Lambda
  }\,, \nonumber\\
  \delta{\cal L}_{{\rm g.k.}}&=&-\ft14\varepsilon^{\mu \nu \rho \sigma }
{\cal G}_{\mu \nu \,\Lambda }\delta {\cal H}_{\rho\sigma}^\Lambda\nonumber\\
&&+\ft18\varepsilon^{\mu \nu \rho \sigma }\Lambda ^Q\left( {\cal
H}_{\mu \nu }^\Lambda X_{Q\Lambda \Sigma }{\cal H}_{\rho\sigma}^\Sigma - 2{\cal
H}_{\mu \nu }^\Lambda X_{Q\Lambda }{}^\Sigma {\cal G}_{\rho \sigma \,\Sigma } -{\cal
G}_{\mu \nu \,\Lambda }X_Q{}^{\Lambda \Sigma }{\cal G}_{\rho \sigma
\,\Sigma }\right) \,,
 \label{LHG}
\end{eqnarray}
where, in the third line, we used the infinitesimal form of
(\ref{rot:calN}):
\begin{eqnarray}
  \label{eq:gauge-var-N-O}
  \delta(\Lambda)\mathcal{N}_{\Lambda\Sigma} &=& \,\Lambda^M\Big[
   -X_{M\Lambda\Sigma} +2\,
  X_{M(\Lambda}{}^\Gamma \mathcal{N}_{\Sigma) \Gamma} +
  \mathcal{N}_{\Lambda\Gamma} \,X_M{}^{\Gamma\Xi}\,
  \mathcal{N}_{\Xi\Sigma} \Big]  \,.
\end{eqnarray}
When we introduce
\begin{equation}
  {\cal G}_{\mu \nu }{}^M= \left({\cal G}_{\mu \nu }{}^\Lambda \,,\,{\cal G}_{\mu \nu \Lambda }\right) \quad \textrm{with} \quad {\cal G}_{\mu \nu }{}^\Lambda \equiv {\cal H}_{\mu \nu
  }{}^\Lambda \,,
 \label{defcalGM}
\end{equation}
we can rewrite the second line of (\ref{LHG}) in a covariant expression,
and when we also use (\ref{Htrafocov}) we get
\begin{eqnarray}
 \delta{\cal L}_{{\rm g.k.}}&=\varepsilon^{\mu \nu \rho \sigma }&\left[
-\ft14{\cal G}_{\mu \nu \,\Lambda }\left(\Lambda^{Q}X_{PQ}{}^\Lambda{\cal
H}_{\rho \sigma }{}^{P}+ X_{(NP)}{}^\Lambda  \Delta B_{\rho \sigma
}{}^{NP}\right)\right.\nonumber\\ &&\left. +\ft18{\cal
G}_{\mu\nu}{}^M{\cal G}_{\rho \sigma }{}^N \Lambda^{Q}X_{QM}{}^R\Omega
_{NR}\right]\,.
 \label{LHG2}
\end{eqnarray}
Clearly, the newly proposed form for ${\cal L}_{\rm g.k.}$ in
(\ref{LgkH}) is still not gauge invariant. This should not come as a
surprise because (\ref{eq:gauge-var-N-O}) contains a constant shift
(i.e., the term proportional to $X_{M\Lambda\Sigma}$), which requires the
addition of extra terms to the Lagrangian as was reviewed in section
\ref{ss:Anomalies} for purely electric gaugings. Also the last term on
the right hand side of (\ref{eq:gauge-var-N-O}) gives extra contributions
that are quadratic in the kinetic function. In the next steps we will see
that besides GCS terms, also terms linear and quadratic in the tensor
field are required to restore gauge invariance. We start with the
discussion of the latter terms.

\subsubsection{Topological terms for the $B$-field and a new constraint}
 \label{ss:topolBconstr}

The second step towards gauge invariance is made by adding topological
terms linear and quadratic in the tensor field $B_{\mu\nu}{}^{NP}$ to the
gauge kinetic term (\ref{LgkH}), namely
\begin{equation}
  \label{eq:LB}
  {\cal L}_{{\rm top},B}= \ft{1}{4}
  \varepsilon^{\mu\nu\rho\sigma}\,
  X_{(NP)}{}^\Lambda \,B_{\mu\nu}{}^{NP} \,
  \Big({\cal F}_{\rho \sigma \,\Lambda }
  +\ft{1}{2}\,X_{(RS)\Lambda} \,B_{\rho \sigma }{}^{RS} \Big)\,.
\end{equation}
Note that for pure electric gaugings $X_{(NP)}{}^\Lambda =0$, as we saw
in (\ref{defXS}). Therefore, in this case this term vanishes, implying
that the tensor fields decouple.

We recall that, up to now, only the closure constraint
(\ref{constraint1}) has been used. We are now going to impose one
\emph{new constraint}:
\begin{equation}
  X_{(NP)}{}^M \Omega _{MQ}  X_{(RS)}{}^Q=0\,.
 \label{newconstraint}
\end{equation}
We will later show that this constraint is implied by the locality
constraint (\ref{constraint2}) and the original representation constraint
of \cite{deWit:2005ub}, i.e. (\ref{repcon}), but also by the locality
constraint and the modified constraint (\ref{relrepcon}) that we
discussed in the introduction. The constraint thus says that
\begin{equation}
  X_{(NP)}{}^\Lambda
  X_{(RS)\Lambda }= X_{(NP)\Lambda }X_{(RS)}{}^\Lambda\,.
 \label{newconstraintexpl}
\end{equation}
A consequence of this constraint that we will use below follows from the
first of (\ref{symmX}) and (\ref{Xsymmcontracted0}):
\begin{equation}
 X_{(PQ)}{}^R D_{MNR}=0\,.
 \label{dcontrsymmX0}
\end{equation}
The variation of $\mathcal{L}_{{\rm top},B}$ is
\begin{eqnarray}
  \label{eq:delta-Ltop}
  \delta\mathcal{L}_{{\rm top},B}& =& \ft{1}{4}  \,
  \varepsilon^{\mu\nu\rho\sigma}X_{(NP)}{}^\Lambda\left[
  {\cal H}_{\mu \nu \Lambda }  \,\delta
 B_{\rho\sigma}{}^{NP} +B_{\rho\sigma}{}^{NP}\delta\mathcal{F}_{\mu\nu\Lambda}
   \right]\\
   &=&  \ft{1}{4}  \,
  \varepsilon^{\mu\nu\rho\sigma} X_{(NP)}{}^\Lambda\left[
  {\cal H}_{\mu \nu \Lambda }   \,\delta
 B_{\rho\sigma}{}^{NP} +2  B_{\rho\sigma}{}^{NP}\left({\cal D}_\mu \delta A_{\nu \Lambda }-
 X_{(RS)\Lambda }A_\mu ^R\delta A_\nu ^S\right)
   \right]\nonumber \,.
\end{eqnarray}

\subsubsection{Generalized Chern-Simons terms}

As in \cite{deWit:2005ub}, we introduce a generalized Chern-Simons term
of the form (these are the last two lines in what they called
$\mathcal{L}_{top}$ in their equation (4.3))
\begin{equation}
  \label{eq:LGCS}
{\cal L}_{\rm GCS} =\varepsilon^{\mu\nu\rho\sigma}A_{\mu}{}^{M} A_{\nu}{}^{N}\left(
\frac{1}{3}\,  X_{MN\,\Lambda}\,\partial_{\rho} A_{\sigma}{}^{\Lambda}
+\frac{1}{6}X_{MN}{}^{\Lambda}\partial_{\rho} A_{\sigma}{}_{\Lambda}
+\frac{1}{8}X_{MN\,\Lambda}X_{PQ}{}^{\Lambda} A_{\rho}{}^{P}A_{\sigma}{}^{Q}\right)\,.
\end{equation}
Modulo total derivatives one can write its variation as (using
(\ref{explicitXalg}) antisymmetrized in $[MNQ]$ and the definition of
$D_{MNP}$ in (\ref{defd}))
\begin{eqnarray}
  \delta{\cal L}_{\rm GCS}& =\varepsilon^{\mu\nu\rho\sigma}&\left[\ft12 {\cal F}_{\mu \nu }{}^\Lambda
  {\cal D}_\rho \delta A_{\sigma\Lambda }-\ft12{\cal F}_{\mu \nu \Lambda }X_{(NP)}{}^\Lambda
  A_\rho{} ^{N}\delta A_{\sigma}{}^P\right.\nonumber\\
  && \left.-D_{MNP}A_\mu {}^M\delta A_\nu {}^N\left(\partial _\rho
  A_\sigma {}^P+\ft38 X_{RS}{}^PA_\rho {}^RA_\sigma {}^S\right) \right]\,.
 \label{delLGCS}
\end{eqnarray}
These variations can be combined with (\ref{eq:delta-Ltop}) to
\begin{eqnarray}
 \delta \left(\mathcal{L}_{{\rm top},B}+ {\cal L}_{\rm GCS}\right) & = & \varepsilon^{\mu\nu\rho\sigma}
 \left[ \ft12  {\cal H}_{\mu \nu }{}^\Lambda
  {\cal D}_\rho \delta A_{\sigma\Lambda }+\ft14 {\cal H}_{\mu \nu \Lambda }X_{(NP)}{}^\Lambda  \left( \delta
 B_{\rho\sigma}{}^{NP}-2
  A_\rho{} ^{N}\delta A_{\sigma}{}^P\right) \right. \nonumber\\
   &   & \left.-D_{MNP}A_\mu {}^M\delta A_\nu {}^N\left(\partial _\rho
  A_\sigma {}^P+\ft38 X_{RS}{}^PA_\rho {}^RA_\sigma {}^S\right) \right]\,.
 \label{totaldelLtop}
\end{eqnarray}

\subsubsection{Variation of the total action}
We are now ready to discuss the symmetry variation of the total
Lagrangian
\begin{equation}
  {\cal L}_{VT}= {\cal L}_{{\rm g.k.}}+{\cal L}_{{\rm top},B}+{\cal L}_{\rm
  GCS}\,,
 \label{LVTsum}
\end{equation}
built from (\ref{LgkH}), (\ref{eq:LB}) and (\ref{eq:LGCS}). We first
check the invariance of (\ref{LVTsum}) with respect to the $\Xi
$-transformations. We see directly from (\ref{LHG2}) that the
gauge-kinetic terms are invariant. The second line of
(\ref{totaldelLtop}) also clearly vanishes inserting (\ref{gaugevar}) and
using (\ref{dcontrsymmX0}). This leaves us with the first line of
(\ref{totaldelLtop}), which, using (\ref{eq:B-transf-1}) and
(\ref{gaugevar}), can be written in a symplectically covariant form:
\begin{equation}
  \delta _\Xi {\cal L}_{VT}=-\ft12 \varepsilon^{\mu\nu\rho\sigma}
 {\cal H}_{\mu \nu }{}^M
 X_{(NP)}{}^Q\Omega _{MQ}  {\cal D}_\rho\Xi _\sigma {}^{NP}\,.
 \label{XitransfLtot}
\end{equation}
The $B$-terms in ${\cal H}$, see (\ref{H}), are proportional to
$X_{(RS)}{}^M$ and thus give a vanishing contribution due to our new
constraint (\ref{newconstraint}). For the ${\cal F}$ terms we can perform
an integration by parts\footnote{Integration by parts with the covariant
derivatives is allowed as (\ref{explicitXalg}) can be read as the
invariance of the tensor $X$ and  (\ref{fromsympl}) as the invariance of
$\Omega $.} and then (\ref{bianchi}) gives again only terms proportional
to $X_{(RS)}{}^M$ leading to the same conclusion. We therefore find that
the $\Xi$-variation of the total action vanishes.

We can thus further restrict to the $\Lambda ^M$ gauge transformations.
According to (\ref{deltaF}), the ${\cal D}_\rho \delta A_{\sigma\Lambda
}$-term in (\ref{totaldelLtop}) can then be replaced by $\ft12\Lambda
^QX_{NQ\Lambda }{\cal H}_{\rho \sigma }{}^N$ (see again footnote
\ref{fn:FH}), which can then be combined with the first term of
(\ref{LHG2}) to form a symplectically covariant expression (the first
term on the right hand side of (\ref{delLVT}) below). Adding also the
remaining terms of (\ref{totaldelLtop}) and (\ref{LHG2}), one obtains,
using (\ref{eq:B-transf-1}),
\begin{eqnarray}
 \delta{\cal L}_{VT}&=\varepsilon^{\mu \nu \rho \sigma }&\left[
\ft14{\cal G}_{\mu \nu  }{}^M\Lambda^{Q}X_{NQ}{}^R\Omega _{MR}{\cal
H}_{\rho \sigma }{}^N +\ft18{\cal G}_{\mu \nu }{}^M{\cal G}_{\rho
\sigma }{}^N \Lambda^{Q}X_{QM}{}^R\Omega _{NR} \right. \nonumber\\
&&+\ft14({\cal H}-{\cal G})_{\mu \nu \,\Lambda } X_{(NP)}{}^\Lambda
\Delta B_{\rho \sigma }{}^{NP}\nonumber\\
   &   & \left.-D_{MNP}A_\mu {}^M{\cal D}_\nu \Lambda ^N\left(\partial _\rho
  A_\sigma {}^P+\ft38 X_{RS}{}^PA_\rho {}^RA_\sigma {}^S\right) \right]\,.
 \label{delLVT}
\end{eqnarray}
We observe that if the ${\cal H}$ in the first line was a ${\cal G}$,
eqs. (\ref{fromsympl}) and (\ref{symmX}) would allow one to write the
first line as an expression proportional to $D_{MNP}$. This leads to the
first line in (\ref{delLVT2}) below.
 The second observation is that the identity $({\cal H}-{\cal
G})^\Lambda =0$ allows one to rewrite the second line of (\ref{delLVT})
in a symplectically covariant way, so that, altogether, we have
\begin{eqnarray}
 \delta{\cal L}_{VT}&=\varepsilon^{\mu \nu \rho \sigma }&\left[
\ft14{\cal G}_{\mu \nu  }{}^M\Lambda^{Q}X_{NQ}{}^R\Omega _{MR}({\cal
H}-{\cal G})_{\rho \sigma }{}^N +\ft38{\cal G}_{\mu \nu }{}^M{\cal G}_{\rho
\sigma }{}^N \Lambda^{Q}D_{QMN} \right. \nonumber\\
&&-\ft14({\cal H}-{\cal G})_{\mu \nu}{}^M\Omega _{MR} X_{(NP)}{}^R
\Delta B_{\rho \sigma }{}^{NP}\nonumber\\
   &   & \left.-D_{MNP}A_\mu {}^M{\cal D}_\nu \Lambda ^N\left(\partial _\rho
  A_\sigma {}^P+\ft38 X_{RS}{}^PA_\rho {}^RA_\sigma {}^S\right) \right]\,.
 \label{delLVT2}
\end{eqnarray}
By choosing
\begin{equation}
  \Delta B_{\rho \sigma }{}^{NP}= -\Lambda ^N {\cal G}_{\rho \sigma }{}^P- \Lambda ^P {\cal
  G}_{\rho \sigma }{}^N\,,
 \label{choiceDeltaB1}
\end{equation}
the result (\ref{delLVT2}) becomes
\begin{eqnarray}
 \delta{\cal L}_{VT}&=\varepsilon^{\mu \nu \rho \sigma }&\left[
\ft38\Lambda^{Q} D_{MNQ}\left( 2{\cal G}_{\mu \nu  }{}^M({\cal H}-{\cal
G})_{\rho \sigma }{}^N+{\cal G}_{\mu \nu }{}^M{\cal G}_{\rho
\sigma }{}^N \right)  \right. \nonumber\\
   &   & \left.-D_{MNP}A_\mu {}^M{\cal D}_\nu \Lambda ^N\left(\partial _\rho
  A_\sigma {}^P+\ft38 X_{RS}{}^PA_\rho {}^RA_\sigma {}^S\right) \right]\,,
 \label{delLVT3}
\end{eqnarray}
which is then proportional to $D_{MNP}$, and hence zero when the original
representation constraint (\ref{constraint3.6}) of \cite{deWit:2005ub} is
imposed.

Our goal is to generalize this for theories with quantum anomalies. These
anomalies depend only on the gauge vectors. The field strengths ${\cal
G}$, (\ref{eq:def-G(H)}), however, also depend on the matrix ${\cal N}$
which itself generically depends on scalar fields. Therefore, we want to
consider modified transformations of the antisymmetric tensors such that
${\cal G}$ does not appear in the final result.

To achieve this, we would like to replace (\ref{choiceDeltaB1}) by a
transformation such that
\begin{equation}
  X_{(NP)}{}^R\Delta B_{\rho \sigma }{}^{NP}=
 -2X_{(NP)}{}^R\Lambda ^N {\cal G}_{\rho \sigma }{}^P
  +\ft32\Omega ^{RM}D_{MNQ}\Lambda ^Q({\cal H}-{\cal G})_{\rho
  \sigma}{}^N\,.
 \label{modDeltaB}
\end{equation}
Indeed, inserting this in (\ref{delLVT2}) would lead to
\begin{eqnarray}
 \delta{\cal L}_{VT}&=\varepsilon^{\mu \nu \rho \sigma }&\left[
\ft38\Lambda^{Q} D_{MNQ}{\cal F}_{\mu \nu }{}^M{\cal F}_{\rho
\sigma }{}^N  \right. \nonumber\\
   &   & \left.-D_{MNP}A_\mu {}^M{\cal D}_\nu \Lambda ^N\left(\partial _\rho
  A_\sigma {}^P+\ft38 X_{RS}{}^PA_\rho {}^RA_\sigma {}^S\right)
  \right]\,,
 \label{delLVTinF}
\end{eqnarray}
where we have used (\ref{dcontrsymmX0}) to delete contributions coming
from the $B_{\mu\nu}{}^{NP}$ term in $\mathcal{H}_{\mu\nu}{}^{M}$ (cf.
(\ref{H})).

The first term on the right hand side of (\ref{modDeltaB}) would follow
from (\ref{choiceDeltaB1}), but the second term cannot in general be
obtained from assigning transformations to $B_{\rho \sigma }{}^{NP}$
(compare with (\ref{symmX})). Indeed, self-consistency of
(\ref{modDeltaB}) requires that the second term on the right hand side be
proportional to $X_{(NP)}{}^R$, which imposes a further constraint on
$D_{MNP}$. We will see in section \ref{ssat} how we can nevertheless
justify the transformation law (\ref{modDeltaB}) by introducing other
antisymmetric tensors. For the moment, we just accept (\ref{modDeltaB})
and explore its consequences.

Expanding (\ref{delLVTinF}) using (\ref{defFmunuX}) and (\ref{gaugevar})
and using a partial integration,  (\ref{delLVTinF}) can be rewritten as
\begin{equation}
  \delta{\cal L}_{VT}=-{\cal A}[\Lambda ]\,,
 \label{finalcancellLVTA}
\end{equation}
where
 \begin{eqnarray}
\mathcal{A}[\Lambda] & = &
-\frac{1}{2}\varepsilon^{\mu\nu\rho\sigma}\Lambda^{P}D_{MNP}\partial_{\mu}A_{\nu}{}^{M}\partial_{\rho}A_{\sigma}{}^{N}\nonumber\\
&&-\frac{1}{4}\varepsilon^{\mu\nu\rho\sigma}\Lambda^{P}\big(D_{MNR}X_{[PS]}{}^{N}+\frac{3}{2}D_{MNP}X_{[RS]}{}^{N}\big)\partial_{\mu}A_{\nu}{}^{M}A_{\rho}{}^{R}A_{\sigma}{}^{S}\,.
\label{finalanomaly}
\end{eqnarray}
This expression formally looks like a symplectically covariant
generalization of the electric consistent anomaly (\ref{gaugeanom}).
Notice, however, that at this point this is really only a formal analogy,
as the tensor $D_{MNP}$ has, a priori, no connection with quantum
anomalies. We will study the meaning of this analogy in more detail in
the next section.  To prove (\ref{finalcancellLVTA}), one uses
(\ref{dcontrsymmX0}) and the preservation of $D_{MNP}$ under gauge
transformations, which follows from preservation of $X$, see
(\ref{explicitXalg}), and of $\Omega $, see (\ref{fromsympl}), and reads
\begin{equation}
  X_{M(N}{}^P\,D_{QR)P}=0\,.
 \label{preservationd}
\end{equation}
For the terms quartic in the gauge fields, one needs the following
consequence of (\ref{preservationd}):
\begin{eqnarray}
(X_{RS}{}^M\,X_{PQ}{}^N\,D_{LMN})_{[RSPL]}&=&-(X_{RS}{}^M\,X_{PM}{}^N\,D_{LQN}+X_{RS}{}^M\,X_{PL}{}^N\,D_{QMN})_{[RSPL]}
\nonumber\\&=&
-(X_{RS}{}^M\,X_{PL}{}^N\,D_{QMN})_{[RSPL]}\,,\label{idsxd}
\end{eqnarray}
where the final line uses (\ref{genJacobi0}) and again
(\ref{dcontrsymmX0}).
\par
Let us summarize the result of our calculation up to the present point.
We have used the action (\ref{LVTsum}) and considered its transformations
under (\ref{gaugevar}) and (\ref{eq:B-transf-1}), where $\Delta B_{\mu
\nu }{}^{NP}$ was undetermined. We used the closure constraint
(\ref{constraint1}) and one new constraint (\ref{newconstraint}). We
showed that the choice (\ref{choiceDeltaB1}) leads to invariance if
$D_{MNP}$ vanishes, which is the representation constraint
(\ref{constraint3.6}) used in the anomaly-free case studied in
\cite{deWit:2005ub}. However, when we use instead the more general
transformation (\ref{modDeltaB}) in the case $D_{MNP}\neq 0$, we obtain
the non-vanishing classical variation (\ref{finalcancellLVTA}). The
corresponding expression (\ref{finalanomaly}) formally looks very similar
to a symplectically covariant generalization of the electric consistent
quantum anomaly.

In order to fully justify and understand this result, we are then left
with the following three open issues, which we will discuss in the
following section:
\begin{enumerate}
  \item The expression (\ref{finalanomaly}) for the non-vanishing classical variation of the action has to
be related to quantum anomalies so that gauge invariance can be restored
at the level of the quantum effective action, in analogy to the electric
case described in section \ref{ss:Anomalies}. This will be done in
section \ref{sssca}.
  \item The meaning of the new constraint (\ref{newconstraint}) that was used to obtain (\ref{finalcancellLVTA}) has to be clarified. This is subject of section \ref{ssnc}.
  \item We have to show how the transformation (\ref{modDeltaB}), which also underlies the result (\ref{finalcancellLVTA}), can be realized. This will be done in section
  \ref{ssat}.
\end{enumerate}



\section{Gauge invariance of the effective action with anomalies}
\label{ss:symmetric}

\subsection{Symplectically covariant anomalies}\label{sssca}

In section \ref{ss:embedding}, we discussed the algebraic constraints
that were imposed on the embedding tensor in ref. \cite{deWit:2005ub} and
that allowed the construction of a gauge invariant Lagrangian with
electric and magnetic gauge potentials as well as tensor fields. Two of
these constraints, (\ref{constraint1}) and (\ref{constraint2}), had a
very clear physical motivation and ensured the closure of the gauge
algebra and the mutual locality of all interacting fields. The physical
origin of the third constraint, the representation constraint,
(\ref{constraint3.6}), on the other hand, remained a bit obscure. In
order to understand its meaning, we specialize it to its purely electric
components,
\begin{equation}
X_{(\Lambda\Sigma\Omega)}=0\,. \label{Xelectric3}
\end{equation}
Given that the components $X_{\Lambda\Sigma\Omega}$ generate axionic
shift symmetries (remember the first term on the right hand side of
(\ref{eq:gauge-var-N-O})), we can identify them with the corresponding
symbols $X_{\Lambda\Sigma\Omega}$ in section \ref{ss:Anomalies}, and
recognize (\ref{Xelectric3}) as the condition for the absence of quantum
anomalies for the electric gauge bosons (see (\ref{Xelectrictotsym})). It
is therefore suggestive to interpret (\ref{constraint3.6}) as the
condition for the absence of quantum anomalies for all gauge fields (i.e.
for the electric and the magnetic gauge fields), and one expects that in
the presence of quantum anomalies, this constraint can be relaxed. We
will show that the relaxation consists in assuming that the symmetric
tensor $D_{MNP}$ defined by (\ref{defd}) is of the form\footnote{The
possibility to impose a relation such as (\ref{newreprconstraint}) is by
no means guaranteed for all types of gauge groups (see e.g.
 \cite{Zagermann:2008gy} for a short discussion in the purely electric case studied in \cite{DeRydt:2007vg}).}
\begin{equation}
D_{MNP} = d_{MNP}\,,\label{newreprconstraint}
\end{equation}
for a symmetric tensor $d_{MNP}$ which describes the quantum gauge
anomalies due to anomalous chiral fermions. In fact, one expects quantum
anomalies from the loops of these fermions, $\psi$, which interact with
the gauge fields via minimal couplings
\begin{eqnarray}
\bar{\psi}\gamma^{\mu}(\partial_{\mu}-
A_{\mu}{}^{\Lambda}\Theta_{\Lambda}{}^{\alpha}\delta _{\alpha}-
A_{\mu\Lambda}\Theta^{\Lambda\alpha}\delta _{\alpha})\psi \,.
\label{coupling}
\end{eqnarray}
Therefore, the anomalies contain -- for each external gauge field (or
gauge parameter) -- an embedding tensor, i.e. $d_{MNP}$ has the following
particular form: 
\begin{equation}
   d_{MNP} = \Theta_M{}^\alpha \Theta_N{}^\beta \Theta_P{}^\gamma d_{\alpha\beta\gamma}\,,\label{dMNPalphabetagamma}
\end{equation}
with $d_{\alpha\beta\gamma}$ being a constant symmetric tensor. In the
familiar context of a theory with a flat scalar manifold, constant
fermionic transformation matrices, $t_{\alpha}$, and the corresponding
minimal couplings, the tensor $d_{MNP}$ is simply proportional to
\begin{equation}
d_{MNP}\propto\Theta_{M}{}^{\alpha}\Theta_{N}{}^{\beta}\Theta_{P}{}^{\gamma}\textrm{Tr}(\{t_{\alpha},t_{\beta}\}t_{\gamma}\},
\label{dcovdef}
\end{equation}
where the trace is over the representation matrices of the
fermions.\footnote{One might wonder how the magnetic vector fields
$A_{\mu\Lambda}$ can give rise to anomalous triangle diagrams, as they
have no propagator due to the lack of a kinetic term. However, it is the
\emph{amputated} diagram with internal fermion lines that one has to
consider.}

We showed that the generalization of the consistent anomaly
(\ref{gaugeanom}) in a symplectically covariant way leads to an
expression of the form (\ref{finalanomaly}) with the $D_{MNP}$-tensor
replaced by $d_{MNP}$. Indeed, the constraint (\ref{newreprconstraint})
implies the cancellation of this quantum gauge anomaly by the classical
gauge variation (\ref{finalcancellLVTA}). Note that it is necessary for
this cancellation that the anomaly tensor $d_{MNP}$ is really constant
(i.e., independent of the scalar fields). We expect this constancy to be
generally true for the same topological reasons that imply the constancy
of $d_{\Lambda \Gamma \Omega}$ in the conventional electric gaugings
\cite{Brandt:1993vd, Brandt:1997au}. In this way we have already
addressed the first issue of the end of the previous section. We are now
going to  show how the constraint (\ref{newreprconstraint}) suffices also
to address the other two issues, (ii) and (iii).

\subsection{The new constraint}\label{ssnc}

We now comment on the constraint (\ref{newconstraint}):
\begin{equation}
  X_{(NP)}{}^M \Omega _{MQ}  X_{(RS)}{}^Q=0\,.
 \label{newconstraintbis}
\end{equation}
We will show that this equation holds if the locality constraint is
satisfied, and (\ref{newreprconstraint}) is imposed on $D_{MNP}$ with
$d_{MNP}$ of the particular form given in (\ref{dMNPalphabetagamma}). To
clarify this, we introduce as in \cite{deWit:2005ub} the `zero mode
tensor'\footnote{Note that the components of $\Omega ^{MN}$ have signs
opposite to those of $\Omega_{MN}$ as given in (\ref{symplstructure}).}
\begin{equation}
    Z^{M\alpha }=\frac12\Omega ^{MN}\Theta _N{}^\alpha \,, \qquad
    \mbox{i.e.}\quad\left\{ \begin{array}{l}
      Z^{\Lambda \alpha }=\ft12 \Theta ^{\Lambda \alpha}\,,\\
      Z_\Lambda {}^\alpha =-\ft12\Theta _\Lambda {}^\alpha \,.
    \end{array}\right.
 \label{defn}
\end{equation}
One then obtains, using (\ref{symmX}), the definition of $X$ in
(\ref{defFmunuX}) and (\ref{dMNPalphabetagamma}) that
\begin{equation}
  X_{(NP)}{}^M= Z^{M\alpha }\Delta_{\alpha NP}\,,
 \label{decompsym}
\end{equation}
for some tensor $\Delta_{\alpha NP}=\Delta_{\alpha PN}$. Due to the fact
that we allow the symmetric tensor $D_{MNP}$ in (\ref{defd}) to be
non-zero and impose the constraint (\ref{newreprconstraint}), this tensor
$\Delta_{\alpha NP}$ is not the analogous quantity called $d_{\alpha MN}$
in \cite{deWit:2005ub}\footnote{We use $\Delta_{\alpha MN}$ in this paper
to denote the analogue (or better: generalization) of what was called
$d_{\alpha MN}$ in \cite{deWit:2005ub}, because $d_{\alpha MN}$ is
reserved in the present paper to denote the quantity
$\Theta_{M}{}^{\beta} \Theta_{N}{}^{\gamma} d_{\alpha \beta \gamma}$ (cf
eq. (\ref{decompdmnr})) related to the quantum anomalies.}, but can be
written as
\begin{equation}
  \Delta_{\alpha NP}= (t_\alpha )_N{}^Q\Omega _{PQ} -3 d_{\alpha \beta \gamma }\Theta
  _N{}^\beta \Theta _P{}^\gamma \,.
 \label{tilded}
\end{equation}
However, the explicit form of this expression will not be relevant. We
will only need that $X_{(NP)}{}^M$ is proportional to $Z^{M\alpha }$.

Now we will finally use the locality constraint (\ref{constraint2}),
which implies
\begin{equation}
  Z^{\Lambda [\alpha }Z_\Lambda {}^{\beta ]}=0\,, \qquad \mbox{i.e.}\qquad
  Z^{M\alpha }Z^{N\beta }\Omega _{MN}=0\,.
 \label{ZZ0}
\end{equation}
This then leads to the desired result (\ref{newconstraintbis}).

The tensor $Z^{M\alpha }$ can be called zero-mode tensor as e.g. the
violation of the usual Jacobi identity (second line of
(\ref{genJacobi0})) is proportional to it. We now show that it also
defines zero modes of $D_{MNR}$. Indeed, another consequence of the
locality constraint is
\begin{equation}
  X_{MN}{}^P\Omega ^{MQ}\Theta _Q^\alpha =0\qquad \rightarrow\qquad  X_{MN}{}^PZ^{M\alpha }=0\,, \qquad
  X_{QM}{}^P\Omega ^{QS}X_{SN}{}^R=0\,.
 \label{fromloc}
\end{equation}
With (\ref{symmX}) and (\ref{Xsymmcontracted0}) this implies
\begin{equation}
  D_{MNR}Z^{R\alpha }=0\,.
 \label{dZ0}
\end{equation}
Note that we did not need (\ref{newreprconstraint}) to achieve this last
result, but that the equation is consistent with it.

\subsection{New antisymmetric tensors} \label{ssat}

Finally, in this section we will justify the transformation
(\ref{modDeltaB}), without requiring further constraints on the
$D$-tensor. That transformation gives an expression for $X_{(NP)}{}^R
\Delta B_{\rho \sigma}{}^{NP}$ that is not obviously a contraction with
the tensor $X_{(NP)}{}^R$ (due to the second term on the right hand side
of (\ref{modDeltaB})). We can therefore in general not assign a
transformation of $B_{\rho \sigma}{}^{NP}$ such that its contraction with
$X_{(NP)}{}^R$ gives (\ref{modDeltaB}). To overcome this problem, we will
have to change the set of independent antisymmetric tensors. The $B
_{\mu\nu } {}^{MN}$ cannot be considered as independent fields in order
to realize (\ref{modDeltaB}). We will, as in \cite{deWit:2005ub},
introduce a new set of independent antisymmetric tensors, denoted by
$B_{\mu\nu\,\alpha}$ for any $\alpha $ denoting a rigid symmetry.

The fields $B_{\mu \nu }{}^{NP}$ and their associated gauge parameters
$\Xi^{NP}$ appeared in the relevant formulae in the form
$X_{(NP)}{}^MB_{\mu \nu }{}^{NP}$ or $X_{(NP)}{}^M\Xi ^{NP}$, see e.g. in
(\ref{gaugevar}), (\ref{deltaF}), (\ref{H}) and (\ref{eq:LB}). With the
form (\ref{decompsym}) that we now have, this can be written as
\begin{equation}
  X_{(NP)}{}^MB_{\mu \nu }{}^{NP}= Z^{M\alpha }\Delta_{\alpha NP}B _{\mu\nu } {}^{NP}\,.
 \label{combinationBinZ}
\end{equation}
We will therefore replace the tensors $B_{\mu\nu}{}^{MN}$ by new tensors
$B_{\mu\nu\,\alpha}$ using
\begin{equation}
 \Delta_{\alpha MN}B _{\mu\nu } {}^{MN} \quad\rightarrow \quad B _{\mu\nu \, \alpha }\,.
 \label{replaceB}
\end{equation}
and consider the $B _{\mu\nu \, \alpha }$ as the independent
antisymmetric tensors. There is thus one tensor for every generator of
the rigid symmetry group. The replacement thus implies that
\begin{equation}
X_{(NP)}{}^MB_{\mu \nu }{}^{NP}\quad \rightarrow \quad  Z^{M\alpha
}B_{\mu \nu\, \alpha }\,.
 \label{translateZB}
\end{equation}
We also introduce a corresponding set of independent gauge parameters
$\Xi_{\mu\,\alpha}$ through the substitution:
\begin{equation}
 \Delta_{\alpha MN}\Xi  _\mu  {}^{MN}\quad\rightarrow \quad \Xi_{\mu \, \alpha }\,.
 \label{defXialpha}
\end{equation}
This allows us to reformulate all the equations in the previous sections
in terms of $B _{\mu\nu \, \alpha}$ and $\Xi_{\mu \, \alpha}$. For
instance we will write:
\begin{eqnarray}
 \delta A_{\mu}{}^{M}  &=&  {\cal
D}_{\mu}\Lambda^{M}-Z^{M\alpha }\Xi_{\mu \, \alpha }\,,\label{gaugevaralpha}\\
 {\cal H}_{\mu\nu}{}^{M} & = & {\cal F}_{\mu\nu}{}^{M}+Z^{M\alpha }B_{\mu \nu\, \alpha }\,, \label{Halpha} \\
  {\cal L}_{{\rm top},B}&=& \ft{1}{4}
  \varepsilon^{\mu\nu\rho\sigma}\,
 Z^{\Lambda \alpha }B_{\mu \nu\, \alpha } \,
  \Big({\cal F}_{\rho \sigma \,\Lambda }
  +\ft{1}{2}\,Z_\Lambda{} ^\beta B_{\rho \sigma\,\beta}\Big)\,. \label{eq:LBalpha}
\end{eqnarray}

We will show that considering $B_{\mu\nu\,\alpha}$ as the independent
variables, we are ready to solve the remaining third issue mentioned at
the end of section \ref{ss:embedding}. To this end, we first note that
all the calculations in section \ref{ss:embedding} remain valid when we
use (\ref{translateZB}) and (\ref{gaugevaralpha})-(\ref{eq:LBalpha}) to
express everything in terms of the new variables $B_{\mu \nu \, \alpha}$
and $\Xi_{\mu \, \alpha}$, because the equations (\ref{newconstraint})
and (\ref{dcontrsymmX0}) we used in section \ref{ss:embedding} are now
simply replaced by (\ref{ZZ0}) and (\ref{dZ0}), respectively.

If we now set, following (\ref{dMNPalphabetagamma}),
\begin{equation}
d_{MNP}  =  \Theta_{M}{}^{\alpha} d_{\alpha NP}\,,\qquad
d_{\alpha NP}= d_{\alpha\beta\gamma} \Theta_N{}^\beta
\Theta_P{}^\gamma\,, \label{decompdmnr}
\end{equation}
then we can define (bearing in mind (\ref{decompsym}))
\begin{eqnarray}
  \label{eq:B-transf-alpha}
   \delta B_{\mu\nu\,\alpha} &=&
   2\,
   {\cal D}_{[\mu} \Xi_{\nu]\alpha} +2 {\Delta}_{\alpha\,NP}A_{[\mu}{}^N \delta
   A_{\nu]}{}^P +\Delta B_{\mu\nu\,\alpha}\,,\nonumber\\
\Delta B_{\mu\nu\,\alpha} &=&
   -2{\Delta}_{\alpha NP}\Lambda^{N}{\cal G}_{\mu\nu}{}^P +3d_{\alpha NP}\Lambda ^N({\cal H}-{\cal G})_{\mu\nu}{}^P\,,
\end{eqnarray}
to reproduce (\ref{modDeltaB}), where the left-hand side of
(\ref{modDeltaB}) is replaced according to (\ref{translateZB}). Here the
covariant derivative is defined as
\begin{equation}
  {\cal D}_{[\mu} \Xi_{\nu]\alpha}=\partial_{[\mu}\Xi_{\nu]\,\alpha}
   + f_{ \alpha\beta  }{}^\gamma  \Theta _P{}^\beta    A_{[\mu}{}^{P}\Xi_{\nu]\,\gamma}\,.
 \label{covderXialpha}
\end{equation}
Of course,  (\ref{eq:B-transf-alpha}) is only fixed modulo terms that
vanish upon contraction with the embedding tensor.

\subsection{Result}

In this section we have seen, so far, that it is possible to relax the
representation constraint (\ref{constraint3.6}) used in ref.
\cite{deWit:2005ub} to the more general condition
(\ref{newreprconstraint}) if one allows for quantum anomalies. The
physical interpretation of the original representation constraint
(\ref{constraint3.6}) of \cite{deWit:2005ub} is thus the absence of
quantum anomalies.

Due to these constraints we obtained the equation (\ref{decompsym}),
which allowed us to introduce the $B_{\mu\nu\,\alpha}$ as independent
variables. All the calculations of section \ref{ss:symplcovgauging} are
then valid with the substitutions given in (\ref{translateZB}) and
(\ref{defXialpha}). We did not impose (\ref{decompsym}) in section
\ref{ss:symplcovgauging}, and therefore we could at that stage only work
with $B_{\mu\nu}{}^{NP}$. However, now we conclude that we need the
$B_{\mu\nu\,\alpha}$ as independent fields and will further only consider
these antisymmetric tensors.

The results of this section can alternatively be viewed as a
covariantization of the results of
\cite{Anastasopoulos:2006cz,DeRydt:2007vg} with respect to
electric/magnetic duality transformations.\footnote{We have not discussed
the complete embedding into $\mathcal{N}=1$ supersymmetry here, which
would include all fermionic terms as well as the supersymmetry
transformations of all the fields. This is beyond the scope of the
present paper.} To further check the consistency of our results, we will
in the next section reduce our treatment to a purely electric gauging and
show that the results of \cite{DeRydt:2007vg} can be reproduced.

\subsection{Purely electric gaugings}

Let us first explicitly write down $D_{MNP}$ in its electric and magnetic
components:
\begin{eqnarray}
D_{\Lambda\Sigma\Gamma} & = & X_{(\Lambda\Sigma\Gamma)}\,,\nonumber\\
3D^{\Lambda}{}_{\Sigma\Gamma} & = & X^{\Lambda}{}_{\Sigma\Gamma}-2 X_{(\Sigma\Gamma)}{}^{\Lambda}\,,\nonumber\\
3
D^{\Lambda\Sigma}{}_{\Gamma}& = & -X_{\Gamma}{}^{\Lambda\Sigma}+2 X^{(\Lambda\Sigma)}{}_{\Gamma} \,,\nonumber\\
D^{\Lambda\Sigma\Gamma} & = & -X^{(\Lambda\Sigma\Gamma)}\,.
\label{componentsd}
\end{eqnarray}

In the case of a purely electric gauging, the only non-vanishing
components of the embedding tensor are electric:
\begin{equation}
  \Theta_M{}^\alpha = \left(\Theta_\Lambda{}^\alpha,\; 0\right)\,.
\end{equation}
Therefore also $X^\Lambda{}_{N}{}^{P} = 0$ and (\ref{dMNPalphabetagamma})
implies that the only non-zero components of $D_{MNP}=d_{MNP}$ are
$D_{\Lambda\Sigma\Omega}$. Therefore, (\ref{componentsd}) reduce to
\begin{equation}
D_{\Lambda\Sigma\Omega} =X_{(\Lambda\Sigma\Omega)}\,,\qquad
X_{(\Sigma\Omega)}{}^\Lambda =0\,,\qquad
 X_\Omega{}^{\Lambda\Sigma} = 0\,.\label{electrreprconstr}
\end{equation}
The non-vanishing entries of the gauge generators are
$X_{\Lambda\Sigma\Gamma}$ and  $X_{\Sigma\Omega}{}^\Lambda=-X_\Sigma
{}^\Lambda {}_\Omega =X_{[\Sigma\Omega]}{}^\Lambda$, the latter
satisfying the Jacobi identities since the right hand side of
(\ref{genJacobi0}) for $MNQR$ all electric indices vanishes. The
$X_{[\Sigma\Omega]}{}^\Lambda$ can be identified with the structure
constants of the gauge group that were introduced e.g. in
(\ref{conditionfinvC1}). The $X_{\Lambda\Sigma\Omega}$ correspond to the
shifts in (\ref{conditionfinvC1}). The first relation in
(\ref{electrreprconstr}) then corresponds to (\ref{Xelectrictotsym}).

The locality constraint is trivially satisfied and the closure relation
reduces to (\ref{closcon}) as expected.

At the level of the action ${\cal L}_{\rm VT}$, all tensor fields drop
out since, when we express everything in terms of the new tensors
$B_{\mu\nu \, \alpha}$, these tensors always appear contracted with a
factor $\Theta^{\Lambda\,\alpha}=0$. In particular, the topological terms
${\cal L}_{{\rm top},B}$ vanish and the modified field strengths for the
electric vector fields ${\cal H}_{\mu\nu}{}^\Lambda$ reduce to ordinary
field strengths:
\begin{eqnarray}
{\cal H}_{\mu\nu}{}^\Lambda&=&2\partial_{[\mu}A_{\nu]}{}^\Lambda +
X_{[\Omega \Sigma]}{}^\Lambda A_\mu{}^\Omega A_\nu{}^\Sigma\,.
\end{eqnarray}
Also the GCS terms (\ref{eq:LGCS}) reduce to their purely electric form
(\ref{SCS1}) with $X_{\Omega \Lambda \Sigma}^{\rm (CS)}=X_{\Omega \Lambda
\Sigma}^{\rm (m)}$. Finally, the gauge variation of ${\cal L}_{\rm VT}$
reduces to minus the ordinary consistent gauge anomaly, as we presented
it in (\ref{gaugeanom}).

This concludes our reinvestigation of the electric gauging with axionic shift symmetries, GCS terms and quantum anomalies as it follows from our more general symplectically covariant treatment. We showed that the more general theory reduces consistently to the known case of a purely electric gauging.

\subsection{On-shell covariance of ${\cal G}_{\mu\nu}{}^M$}

For completeness, we will show in this section that ${\cal
G}_{\mu\nu}{}^M$ (as defined in (\ref{eq:def-G(H)}) and (\ref{defcalGM}))
is the object that transforms covariantly on-shell, rather than ${\cal
H}_{\mu\nu}{}^M$. We consider the total action (\ref{LVTsum}), where now
${\cal L}_{{\rm top},B}$ is given by (\ref{eq:LBalpha}), and in ${\cal
L}_{{\rm g.k.}}$, the expression (\ref{Halpha}) is used. We write the
general variation of this action under generic variations $\delta
A_\mu{}^M,\,\delta B_{\mu\nu\,\alpha}$ of
$A_\mu{}^M,\,B_{\mu\nu\,\alpha}$. The variation of ${\cal L}_{{\rm
g.k.}}$ has a contribution only from $\mathcal{H}^\Lambda$, since the
matrix $\mathcal{N}$ is inert under variations of $A_\mu{}^M$ and
$B_{\mu\nu\,\alpha}$, and thus will be given by the first term in the
expression of $\delta {\cal L}_{{\rm g.k.}}$ in (\ref{LHG}). Summing this
variation with the variation of the topological terms
(\ref{totaldelLtop}) we find:
\begin{eqnarray}
\delta {\cal L}_{VT}&=&\varepsilon^{\mu\nu\rho\sigma}
 \left[ \ft12  {\cal G}_{\mu \nu }{}^M
  {\cal D}_\rho \delta A_{\sigma}{}^N\,\Omega_{MN}\right.\nonumber\\
&&\phantom{\varepsilon^{\mu\nu\rho\sigma}}\left.  +\ft14 \,({\cal H}_{\mu
\nu \Lambda }-{\cal G}_{\mu \nu \Lambda })\, \left(
Z^{\Lambda\alpha}\delta B_{\rho\sigma\,\alpha}-2X_{(NP)}{}^\Lambda
  A_\rho{} ^{N}\delta A_{\sigma}{}^P\right) \right. \nonumber\\
   &   & \phantom{\varepsilon^{\mu\nu\rho\sigma}}\left.-D_{MNP}A_\mu {}^M\delta A_\nu {}^N\left(\partial _\rho
  A_\sigma {}^P+\ft38 X_{RS}{}^PA_\rho {}^RA_\sigma {}^S\right) \right]\,.\label{delLVTsum}
\end{eqnarray}
This allows us to determine the equations of motion for the independent
tensor fields $B_{\mu\nu\, \alpha}$:
\begin{equation}
\frac{\delta {\cal L}_{VT}}{\delta B_{\mu\nu \,\alpha}}\approx 0\qquad
\Leftrightarrow\qquad ({\cal H}-{\cal G})_{\mu \nu \Lambda
}\,Z^{\Lambda \alpha}=\ft12({\cal H}-{\cal G})_{\mu \nu \Lambda
}\,\Theta^{\Lambda \alpha}\approx 0\,,\label{eqb}
\end{equation}
which tells us that the equations of motion
imply\footnote{Identifications on shell are indicated by $\approx $.}
that just some ${\cal H}_{\mu\nu \, \Lambda}$ are identified on-shell
with the corresponding ${\cal G}_{\mu\nu \Lambda}$. More precisely, these
are the tensors ${\cal H}_{\mu\nu \, \Lambda}$ that are singled out by
the contraction with $\Theta^{\Lambda \alpha}$; they thus correspond to
those magnetic vectors $A_{\mu \Lambda}$ that enter the action. From
(\ref{eqb}), together with the constraint (\ref{newreprconstraint}) and
the particular form (\ref{dMNPalphabetagamma}) for $d_{MNP}$, we also see
that
\begin{eqnarray}
({\cal H}_{\mu \nu }{}^P-{\cal G}_{\mu \nu
}{}^P)\,D_{PMN}\approx 0\,.\label{GHeqs}
\end{eqnarray}
The properties (\ref{eqb}) and (\ref{GHeqs}) will be used next to prove
that the tensor which is actually on-shell covariant under gauge-induced
duality transformations is ${\cal G}_{\mu \nu }{}^M$ and not ${\cal
H}_{\mu \nu }{}^M$.

\par Given the complete gauge variation for
the antisymmetric tensor fields (\ref{eq:B-transf-alpha}),  we can write
down the explicit gauge transformation properties of
$\mathcal{H}_{\mu\nu}{}^{M}$ and $\mathcal{G}_{\mu\nu}{}^{M}$, which
generalize those found in \cite{deWit:2005ub,deWit:2007mt} for
$D_{MNP}=0$:
\begin{eqnarray}
\delta \mathcal{H}_{\mu\nu}{}^{M}&=&-\Lambda^Q\,X_{QP}{}^M\,
\mathcal{H}_{\mu\nu}{}^{P}+\Lambda^Q\,\left(2\,X_{(QP)}{}^M+
\ft{3}{2}\,\Omega^{MN}\,D_{NPQ}\right)\,(\mathcal{H}_{\mu\nu}{}^{P}-\mathcal{G}_{\mu\nu}{}^{P})\,,\nonumber\\
\delta \mathcal{G}_{\mu\nu}{}^{\Lambda}&=&-\Lambda^Q\,X_{QP}{}^\Lambda\,
\mathcal{G}_{\mu\nu}{}^{P}+\Lambda^Q\,\hat{X}_{PQ}{}^\Lambda\,(\mathcal{H}_{\mu\nu}{}^{P}-\mathcal{G}_{\mu\nu}{}^{P})\,,\nonumber\\
\delta \mathcal{G}_{\mu\nu\Lambda}&=&-\Lambda^Q\,X_{QP\Lambda}
\mathcal{G}_{\mu\nu}{}^{P}+\frac{1}{2}\,\varepsilon_{\mu\nu\rho\sigma}\,\mathcal{I}_{\Lambda\Sigma}\,\Lambda^Q\,\hat{X}_{PQ}{}^{\Sigma}\,
(\mathcal{H}^{\rho\sigma P}-\mathcal{G}^{\rho\sigma
P})\nonumber\\&&+\mathcal{R}_{\Lambda\Sigma}\,\Lambda^Q\,\hat{X}_{PQ}{}^{\Sigma}\,(\mathcal{H}_{\mu\nu}{}^{P}-\mathcal{G}_{\mu\nu}{}^{P})\,,
\label{delfullHG}
\end{eqnarray}
where we have used the following short-hand notation:
\begin{eqnarray}
\hat{X}_{PQ}{}^M&\equiv& X_{PQ}{}^M+\ft{3}{2}\,\Omega^{MN}\,D_{NPQ}\,.
\end{eqnarray}
The first line of (\ref{delfullHG}) follows from (\ref{Htrafocov}) and
(\ref{modDeltaB}). The second transformation is a component of the first
one since ${\cal G}_{\mu\nu}{}^{\Lambda}={\cal H}_{\mu\nu}{}^{\Lambda}$,
and for the transformation of $\mathcal{G}_{\mu\nu\Lambda}$ we use
(\ref{eq:gauge-var-N-O}).

{}From (\ref{eqb}) and (\ref{GHeqs}) we see that, on-shell, the terms
containing
$(\mathcal{H}_{\mu\nu}{}^{P}-\mathcal{G}_{\mu\nu}{}^{P})\,\hat{X}_{PQ}{}^M$
vanish. Therefore we conclude that, as opposed to
$\mathcal{H}_{\mu\nu}{}^{M}$, the tensor $\mathcal{G}_{\mu\nu}{}^{M}$ is
on-shell gauge covariant and the gauge algebra closes on it modulo field
equations. Consistency of course requires that field equations transform
into field equations, and indeed it can be shown that:
\begin{eqnarray}
\delta
(\mathcal{H}_{\mu\nu\Lambda}-\mathcal{G}_{\mu\nu\Lambda})&=&\Lambda^Q\,\left(\hat{X}_{PQ\Lambda}+\mathcal{R}_{\Lambda\Sigma}\,\hat{X}_{PQ}{}^\Sigma\right)\,
(\mathcal{H}_{\mu\nu}{}^{P}-\mathcal{G}_{\mu\nu}{}^{P})\nonumber\\&&+\ft{1}{2}\,\varepsilon_{\mu\nu\rho\sigma}\,\mathcal{I}_{\Lambda\Sigma}\,\Lambda^Q\,\hat{X}_{PQ}{}^{\Sigma}\,
(\mathcal{H}^{\rho\sigma P}-\mathcal{G}^{\rho\sigma P})\,.
\end{eqnarray}




\section{A simple nontrivial example \label{ss:example}}

Let us now briefly illustrate the above results by means of a simple
example. We consider a theory with a rigid symmetry group embedded in the
electric/magnetic duality group $Sp(2,\mathbb{R})$. The embedding in the
symplectic transformations is given by
\begin{equation}\label{embedding}
  t_{1M}{}^N = \left(\begin{array}{cc}
    1&0\\0&-1
  \end{array}\right)\,,\qquad t_{2M}{}^N = \left(\begin{array}{cc}
    0&0\\1&0
  \end{array}\right)\,,\qquad t_{3M}{}^N = \left(\begin{array}{cc}
    0&1\\0&0
  \end{array}\right)\,,
\end{equation}
i.e. $t_2{}^{11}=1$. Let us consider the following subset of duality
transformations:
\begin{equation}
  {\cal S}^M{}_N{}=\delta^M{}_N -\Lambda^P X_{PN}{}^M\,,
  \quad \mbox{with generators}\quad
  X_{PM}{}^N = \left(\begin{array}{cc}
    0&0\\X_P{}^{11}&0
  \end{array}\right)\,,
\end{equation}
where $\Lambda^P$ is the rigid transformation parameter.  The tensor $X$
is related to the embedding of the symmetries in the symplectic algebra
using the embedding tensor,
\begin{equation}
  X_{PM}{}^N = \sum_{\alpha = 1}^3 \Theta_P{}^\alpha t_{\alpha M}{}^N\,.
 \label{Xsumalpha}
\end{equation}
We have thus chosen the embedding tensor
\begin{equation}
 \Theta_P{}^1 = 0\,,\qquad  \Theta_P{}^2 = X_P{}^{11}\,,\qquad  \Theta_P{}^3 = 0\,.
\end{equation}

We now want to promote ${\cal S}^M{}_N$ to be a gauge transformation,
i.e., we take the $\Lambda^N=\Lambda^N(x)$ spacetime dependent and the
$X_{PM}{}^N$ are the gauge generators. This obviously corresponds to a
magnetic gauging, as (\ref{electrreprconstr}) is violated, and therefore
requires the formalism that was developed in \cite{deWit:2005ub} and
reviewed in section \ref{ss:symplcovgauging}. The locality constraint
(\ref{constraint2}) is automatically satisfied, as only the index value
$\alpha =2$ appears, and closure of the gauge algebra spanned by the
$X_{PM}{}^N$ requires that we impose (\ref{constraint1}), where only the
right-hand side is non-trivial. It requires $\Theta _1{}^2=0$, and thus
the only gauge generators that are consistent with this constraint are
 \begin{equation}
  X_{PM}{}^N = (X_{1M}{}^N, \; X^1{}_M{}^N)\,, \quad {\rm with}\quad X_{1M}{}^N = 0\,,\quad  X^1{}_M{}^N = \left(\begin{array}{cc}
    0&0\\X^{111}&0
  \end{array}\right)\label{genexample}\,.
\end{equation}
Note that this choice still violates the original linear representation
constraint (\ref{constraint3.6}), as (\ref{componentsd}) gives
$D^{111}=-X^{111}\neq 0$. However, as we saw in section
\ref{ss:embedding}, this does not prevent us from performing the gauging
with generators $X_{PM}{}^N$ given in (\ref{genexample}). We introduce a
vector $A_\mu{}^M$ which contains an electric and a magnetic part,
$A_\mu{}^1$ and $A_{\mu 1}$. Note that only the magnetic vector couples
to matter via covariant derivatives since the embedding tensor projects
out the electric part. In what follows, we also assume the presence of
anomalous couplings between the magnetic vector and chiral fermions. As
we will now review, this justifies the nonzero $X^{111}\neq 0$, since it
will give rise to anomaly cancellation terms in the classical gauge
variation of the action. More precisely, we will have to require that
\begin{equation}
  \Theta ^{12}= X^{111}\,,\qquad -X^{111} = d^{111} = (X^{111})^3 {\tilde
  d}_{222}\,,
 \label{exampledabc}
\end{equation}
where we introduced ${\tilde d}_{222}$ as the component of
$d_{\alpha\beta\gamma}$.

To show this, we first introduce a kinetic term for the electric vector
fields:
\begin{equation}
  {\cal L}_{\rm g.k.} = \ft{1}{4}\; e\; {\cal I} \;{\cal H}_{\mu\nu}{}^1{\cal H}^{\mu\nu \,1}
  -\ft{1}{8}\;{\cal R} \;\varepsilon^{\mu\nu\rho\sigma}  {\cal H}_{\mu\nu}{}^1{\cal H}_{\rho\sigma}{}^1,
\end{equation}
where we introduced the modified field strength  (\ref{Halpha})
\begin{equation}
  {\cal H}_{\mu\nu}{}^1 = 2\partial_{[\mu} A_{\nu]}{}^1 + \frac{1}{2}X^{111}B_{\mu\nu 2}\,,
\end{equation}
which depends on a tensor field $B_{\mu\nu 2}$ and therefore transforms
covariantly under
\begin{eqnarray}
  \label{deltaA}\delta A_\mu{}^1 &=& \partial_\mu \Lambda^1 + X^{111}A_{\mu\,1} \Lambda_1 - \frac{1}{2}X^{111}\Xi_{\mu 2}\,,\nonumber\\
   \delta B_{\mu\nu 2} &=& 2 \partial_{[\mu}\Xi_{\nu]2}+4A_{[\mu\,1}\partial_{\nu]}\Lambda_1-6 \Lambda_1
   \partial_{[\mu}A_{\nu]\,1}-\Lambda _1{\cal G}_{\mu \nu \,1}\,,\nonumber\\
   \delta A_{\mu 1}&=&\partial _\mu \Lambda _1\,.
\end{eqnarray}
This follows from (\ref{eq:B-transf-alpha}) since the only nonzero
component of $\Delta_{2MN}$ is $\Delta_2{}^{11}=2$ and for $d_{2MN}$ we
have only $d_2{}^{11}=-1$. One can check that
\begin{eqnarray}
  \delta {\cal H}_{\mu\nu}{}^1 &=& -\ft12 X^{111} \Lambda_1({\cal H}+{\cal G})_{\mu\nu\,1}\,, \quad \mbox{with} \nonumber\\
 && {\cal H}_{\mu\nu\,1}={\cal F}_{\mu\nu\,1}=2 \partial_{[\mu}A_{\nu]1}\,,\qquad {\cal G}_{\mu\nu\,1} \equiv {\cal R}{\cal H}_{\mu\nu}{}^1 +\frac{1}{2}e {\cal I}\varepsilon_{\mu\nu\rho\sigma}{\cal H}^{\rho\sigma\,1}\,.
\end{eqnarray}
Under gauge variations, the real and imaginary part of the kinetic
function transform as follows (cf. (\ref{eq:gauge-var-N-O})):
\begin{equation}
  \delta {\cal I} = 2\Lambda_1 X^{111} {\cal R}{\cal I}\,,\qquad \delta {\cal R} = \Lambda_1 X^{111} \left({\cal R}^2-{\cal I}^2\right)\,.
\end{equation}
 Then it's a short calculation to show that
 \begin{equation}
   \label{deltaLgkex}\delta {\cal L}_{\rm g.k.} = \ft{1}{4}  \varepsilon^{\mu\nu\rho\sigma} \Lambda_1 X^{111} {\cal G}_{\mu\nu\,1}\partial  _\rho A_{\sigma 1} \,.
 \end{equation}
This is consistent with (\ref{LHG2}).

In a second step, we add the topological term (\ref{eq:LBalpha})
\begin{equation}
   {\cal L}_{{\rm top},B}=\ft{1}{4} \varepsilon^{\mu\nu\rho\sigma} X^{111} B_{\mu\nu 2}\partial_{[\rho} A_{\sigma]\,1}\,.
\end{equation}
The gauge variation of this term is equal to (up to a total derivative)
\begin{equation}
  \label{deltatopex}\delta {\cal L}_{{\rm top},B}=-\ft14\Lambda_1 X^{111} \varepsilon^{\mu\nu\rho\sigma} \left(\partial_{\mu}A_{\nu\,1}\right)\left(2\partial_{\rho}A_{\sigma\,1}+{\cal G}_{\rho \sigma \,1}\right)\,.
\end{equation}
The generalized Chern-Simons term (\ref{eq:LGCS}) vanishes in this case.
Combining (\ref{deltaLgkex}) and (\ref{deltatopex}), one derives
\begin{equation}
  \label{deltaLex}
  \delta\left({\cal L}_{\rm g.k.}+{\cal L}_{{\rm top},B}\right)
 = -\ft12\Lambda_1 X^{111} \left(\partial_{\mu}A_{\nu\,1}\right)\left(\partial_{\rho}A_{\sigma\,1}\right)\varepsilon^{\mu\nu\rho\sigma}\,.
\end{equation}
This cancels the magnetic gauge anomaly whose form can be derived from
(\ref{finalanomaly}),
\begin{eqnarray}
  {\cal A}[\Lambda]&=&-\ft{1}{2} \varepsilon^{\mu\nu\rho\sigma}\Lambda_1 d^{111} \left(\partial_\mu A_{\nu\,1}\right) \left(\partial_\rho A_{\sigma\,1}\right)\,,
 \label{examplean}
\end{eqnarray}
if we remember that $X^{111}=-D^{111}=-d^{111}$. Note that the electric
gauge fields do not appear which corresponds to the fact that the
electric gauge fields do not couple to the chiral fermions.

A simple fermionic spectrum that could yield such an anomaly
(\ref{examplean}) is given by, e.g., three chiral fermions with canonical
kinetic terms and quantum numbers $Q=(-1),(-1),(+2)$ under the $U(1)$
gauged by $A_{\mu\, 1}$. Indeed, with this spectrum, we would have
$\trace(Q)=0$, i.e., vanishing gravitational anomaly, but a cubic Abelian
gauge anomaly $d^{111} \propto \trace(Q^3)= +6$.


\section{Conclusions}
\label{ss:Conclusions}
In this paper we have shown how general gauge theories with axionic shift symmetries, generalized Chern-Simons terms and quantum anomalies \cite{DeRydt:2007vg} can be formulated in a way that is covariant with respect to electric/magnetic duality transformations.
This generalizes previous work of \cite{deWit:2005ub}, in which only \emph{classically} gauge invariant theories with anomaly-free fermionic spectra were considered. Whereas the work \cite{deWit:2005ub} was modelling extended (and hence automatically anomaly-free) gauged supergravity theories, our results here can be applied to general $\mathcal{N}=1$ gauged supergravity theories with possibly anomalous fermionic spectra. Such anomalous fermionic spectra are a natural feature of many string
compactifications, notably of intersecting brane models in type II orientifold compactifications, where also GCS terms  frequently occur \cite{Anastasopoulos:2006cz}. Especially in combination with background fluxes, such compactifications may
naturally lead to 4D actions with tensor fields and gaugings in unusual duality frames. Our formulation  accommodates all these non-standard formulations, just as ref. \cite{deWit:2005ub} does in the anomaly-free case.

At a technical level, our results were obtained by relaxing the so-called
representation constraint to allow for a symmetric three-tensor $d_{MNP}$
that parameterizes the quantum anomaly. In contrast to the other
constraints for the embedding tensor, this modified representation
constraint is not homogeneous in the embedding tensor, which is a novel
feature in this formalism. Also our treatment gave an interpretation for
the physical meaning of the ``representation'' constraint: In its
original form used in \cite{deWit:2005ub}, it simply states the absence
of quantum anomalies. It is interesting, but in retrospect not
surprising, that the extended supergravity theories from which the
original constraint has been derived in \cite{deWit:2005ub}, need this
constraint for their internal classical consistency.

It would be interesting to embed our results in a manifestly
supersymmetric framework. Likewise, it would be interesting to study
explicit $\mathcal{N}=1$ string compactifications within the framework
used in this paper, making use of manifest duality invariances. Another
topic we have not touched upon are K{\"a}hler anomalies
\cite{Freedman:1976uk,Chamseddine:1995gb,Castano:1995ci,LopesCardoso:1991zt,
LopesCardoso:1992yd,Derendinger:1991kr,Derendinger:1991hq,Kaplunovsky:1994fg,
Kaplunovsky:1995jw,Freedman:2005up,Elvang:2006jk} in $\mathcal{N}=1$
supergravity or gravitational anomalies. We hope to return to some of
these questions in the future.


\medskip
\section*{Acknowledgments.}

\noindent We are grateful to Bernard de Wit and Henning Samtleben  for
useful discussions. This work is supported in part by the European
Community's Human Potential Programme under contract MRTN-CT-2004-005104
`Constituents, fundamental forces and symmetries of the universe'. The
work of J.D.R. and A.V.P. is supported in part by the FWO - Vlaanderen,
project G.0235.05 and by the Federal Office for Scientific, Technical and
Cultural Affairs through the `Interuniversity Attraction Poles Programme
-- Belgian Science Policy' P6/11-P. The work of J.D.R. has also been
supported by a Marie Curie Early Stage Research Training Fellowship of
the European Community's Sixth Framework Programme under contract number
(MEST-CT-2005-020238-EUROTHEPHY). The work of T.S. and M.Z. is supported
by the German Research Foundation (DFG) within the Emmy-Noether Programme
(Grant number  ZA 279/1-2).

\newpage


\providecommand{\href}[2]{#2}\begingroup\raggedright\endgroup

\end{document}